\documentclass[aps,prd,twocolumn,showpacs,preprintnumbers]{revtex4}

\usepackage{graphicx}
\usepackage{epsfig}

\def\theta{\vartheta}

\newcommand{\be}{\begin{equation}}
\newcommand{\ee}{\end{equation}}
\newcommand{\ba}{\begin{eqnarray}}
\newcommand{\ea}{\end{eqnarray}}
\newcommand{\lsim}   {\mathrel{\mathop{\kern 0pt \rlap
  {\raise.2ex\hbox{$<$}}}
  \lower.9ex\hbox{\kern-.190em $\sim$}}}
\newcommand{\gsim}   {\mathrel{\mathop{\kern 0pt \rlap
  {\raise.2ex\hbox{$>$}}}
  \lower.9ex\hbox{\kern-.190em $\sim$}}}

\begin{document}

\preprint{}

\title{Exploiting the neutronization burst of a galactic supernova}

\author{M.~Kachelrie\ss\ and R.~Tom\`as}

\affiliation{Max-Planck-Institut f\"ur Physik (Werner-Heisenberg-Institut),
F\"ohringer Ring 6, D--80805 M\"unchen} 

\author{R.~Buras, H.-Th.~Janka, A.~Marek, and M.~Rampp}

\affiliation{Max-Planck-Institut f\"ur Astrophysik,
  Karl-Schwarzschild-Str. 1, D--85741 Garching}

\date{December 3, 2004} 

\begin{abstract}
One of the robust features found in simulations of core-collapse
supernovae (SNe) is the prompt neutronization burst, i.e. the first
$\sim 25$~milliseconds after bounce when the SN emits with very high
luminosity mainly $\nu_e$ neutrinos. We examine the dependence of this
burst on variations in the input of current SN models and find that
recent improvements of the electron capture rates as well as 
uncertainties in the nuclear equation of state or a 
variation of the progenitor mass have only little effect on the 
signature of the neutronization peak in a
megaton water Cherenkov detector for different neutrino mixing schemes. 
We show that exploiting the time-structure of the neutronization peak
allows one to identify the case of a normal mass hierarchy and
large 13-mixing angle $\theta_{13}$, where the peak is absent. 
The robustness of the predicted total event number in the
neutronization burst makes a measurement of the distance to the SN
feasible with a precision of about 5\%, even in the likely case that
the SN is optically obscured.  
\end{abstract}


\maketitle

\section{Introduction}

Despite the enormous progress of neutrino physics in the last decade, 
many open questions remain to be solved. Among them are two, the 
mass hierarchy---normal versus inverted mass spectrum---and the
value of the 13-mixing angle $\theta_{13}$, where the observation of
neutrinos from a  core-collapse supernova (SN) could provide
important clues~\cite{Dighe:1999bi,Lunardini:2003eh,Takahashi:2003bj}. 
The neutrino emission by a SN can be divided schematically into four
stages: Infall phase, neutronization burst, accretion phase, 
and Kelvin-Helmholtz
cooling phase. The bulk of SN neutrinos are emitted  in all flavors
during the last two phases with small differences between the
$\bar\nu_e$ and $\bar\nu_{\mu,\tau}$ 
spectra~\cite{Raffelt2001,Buras2003,Keil2003}.
Moreover, the absolute values of the average neutrino energies as well
as the relative size of the luminosities during the accretion and
cooling phases are not known with sufficient precision. As a
consequence, a straightforward extraction of oscillation parameters
from the SN neutrino signal during the accretion and cooling phase
seems hopeless.

An alternative is the use of observables that do not rely on SN
parameters. Such observables require only that the initial neutrino
fluxes $F_i^0$ are different functions of energy and time, 
$F^0_{\nu_e,\bar\nu_e}(E,t) \neq F_{\nu_x,\bar\nu_x}^0(E,t)$, 
where $x=\{ \mu,~\tau\}$. 
Since the interaction of neutrinos with matter depends on their flavor, 
identical energies and luminosities for $\nu_e$, $\bar\nu_e$ and
$\nu_x$ would require a conspiracy of interaction rates and chemical
composition inside the neutrinospheres. 
Examples for such observables are the modulations in the SN
neutrino signal caused by the passage of the neutrinos through the
Earth~\cite{earth} or by the propagation of shock waves  through the SN
envelope~\cite{shock,Tomas:2004gr}. If the mixing
angle $\theta_{13}$ is known to be large, $\sin^2\theta_{13}\gsim 10^{-3}$,
an observation of Earth matter or shock wave effects in the
experimentally most important $\bar\nu_e$ channel would imply a normal
or inverted mass hierarchy,  
respectively. If however the value of $\theta_{13}$ is not known, a
degeneracy exists between the case of a normal mass hierarchy and
large $\theta_{13}$ (scenario A, cf. table~1) and the case of small
$\theta_{13}$, $\sin^2\theta_{13}\lsim 10^{-5}$, and any hierarchy 
(scenario C): scenario A and C both predict the same $\bar\nu_e$
signature in a water Cherenkov detector. 

A different way to extract reliable information about neutrino mixing
parameters is to use characteristics in the neutrino emission of SNe that are
model independent. One of the most robust features of  numerical SN
simulations is the so-called neutronization $\nu_e$ burst~\cite{burst},
which takes place during the first $\sim 25$~ms after the core
bounce. The small number of events expected during this time period
is compensated by the moderate dependence of the $\nu_e$ burst on physical
parameters like the progenitor mass or details of the SN models.
In Sec.~II, we discuss the astrophysical aspects of
this burst in detail, emphasizing the robustness of the neutrino
luminosities against variations in the input of the SN models.
In Sec.~III, we study the signature of the neutronization peak in a
megaton water Cherenkov detector for different neutrino mixing
schemes. We argue that exploiting the time-structure of the
neutronization peak allows one to identify the neutronization
burst even if the SN is not visible in the optical. 
A non-observation of the
neutronization burst identifies the case of a normal mass hierarchy
and large 13-mixing angle $\theta_{13}$ (case A), thus breaking the
degeneracy between the neutrino mixing scenarios A and C.
Moreover we find that for a given neutrino mixing scenario the
systematic uncertainty due to unknown SN parameters affects 
only little the total number of events in the neutronization burst. 
As we discuss in Sec.~IV, this robustness of the theoretical
prediction makes a measurement of the distance to a SN located at 10
kpc feasible with a precision of about 5\%. Such an accuracy is
comparable to optical methods using the SN light curve, which have an error
between 5 and 10\%. If the SN is optically obscured, measuring its
distance through the $\nu_e$ burst is crucial for
estimating the total binding energy released or to limit the strength 
of the gravitational wave signal~\cite{Mueller:2003fs} emitted by the SN. 
Finally, we summarize our results in Sec.~V.

\section{Astrophysical aspects of the neutronization burst}
\label{astro}

\begin{figure*}[!]
\epsfig{file=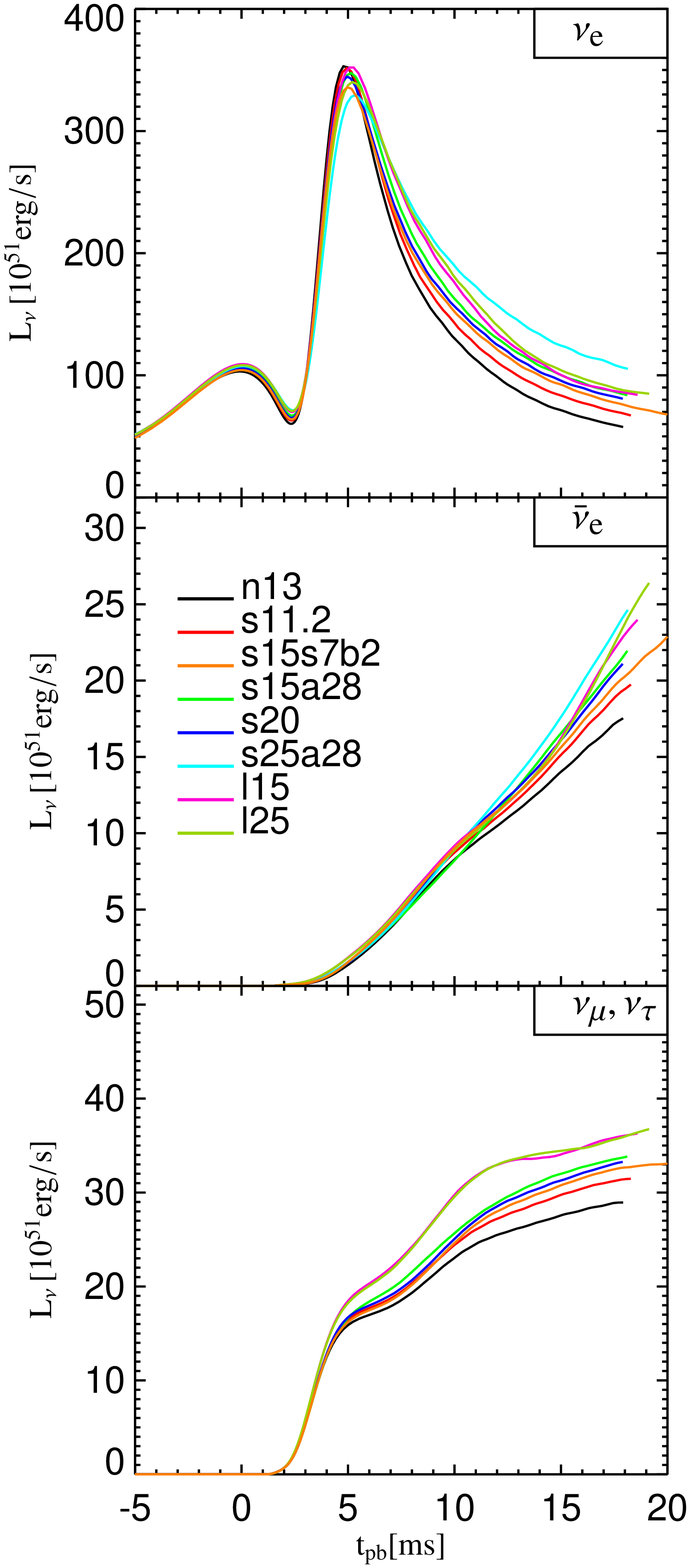,width=0.32\hsize,angle=0}
\epsfig{file=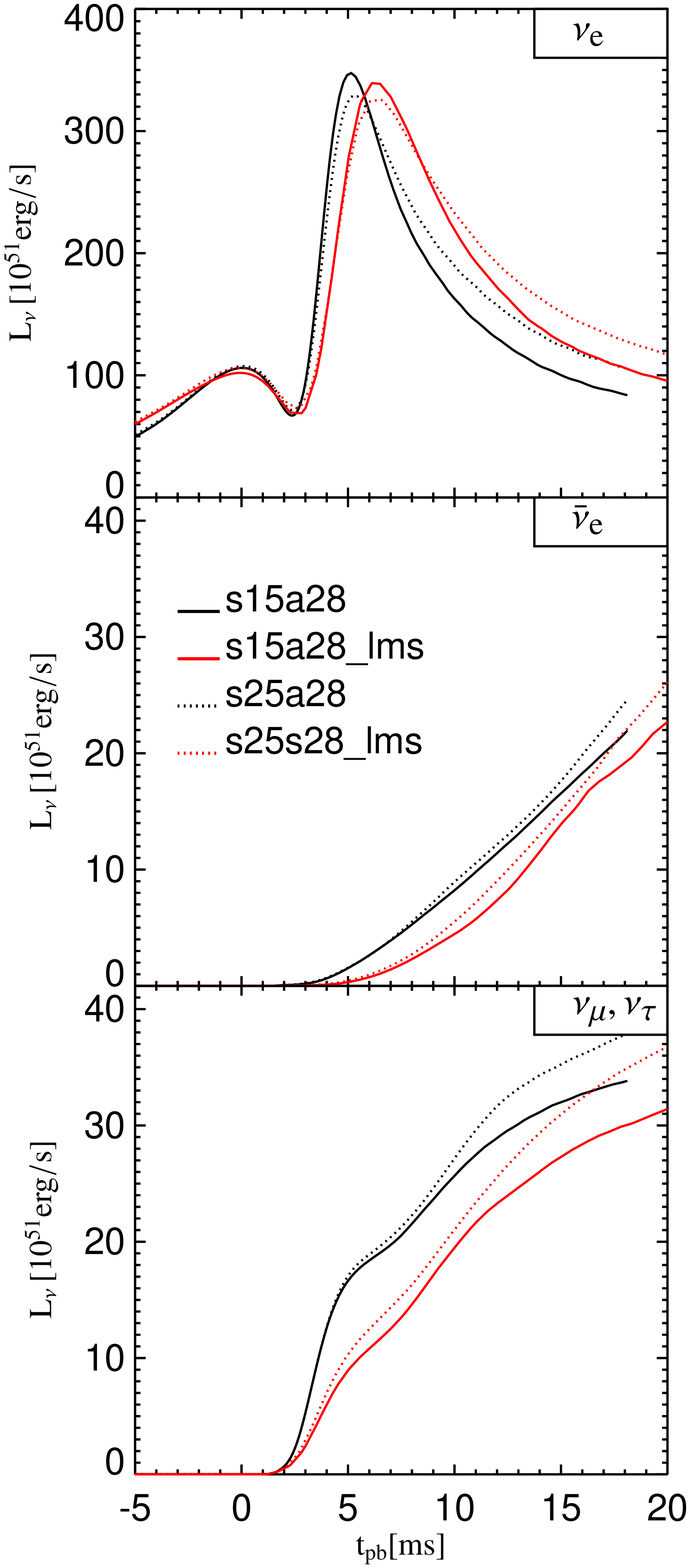,width=0.32\hsize,angle=0}
\epsfig{file=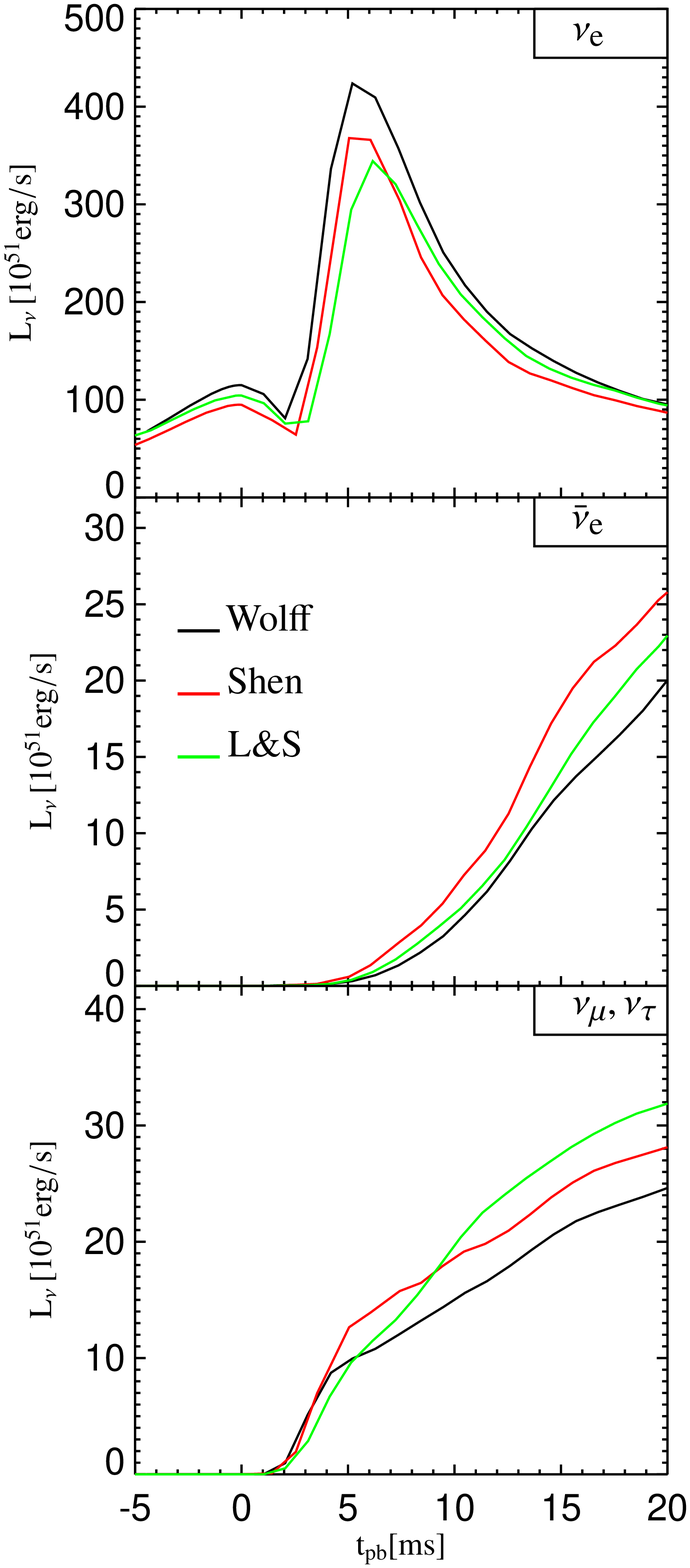,width=0.32\hsize,angle=0}
\caption{\label{lumi}
Luminosities as functions of time for $\nu_e$ (top),
$\bar\nu_e$ (middle) and heavy-lepton neutrinos (bottom).
In the left column results for 
different progenitor stars between 11.2$\,M_{\odot}$
and 25$\,M_{\odot}$ (left column; the progenitor mass is
indicated by the number after the first letter of the model 
name~\cite{progenitors}) are shown, in the middle
column for simulations with the new treatment of electron
captures by nuclei during stellar core collapse according to
Langanke, Mart\'{\i}nez-Pinedo and Sampaio (LMS;
red solid and dotted
lines) compared to the traditional description (black lines)
in case of a 15$\,M_{\odot}$ and a 25$\,M_{\odot}$ star.
The right column shows results for three different nuclear
equations of state applied to the collapse of a 15$\,M_{\odot}$
progenitor (see text for more details).
The luminosities are given for an observer at rest, evaluated
at a radius of 400$\,$km with a corresponding
time retardation of about 1$\,$ms.
Time is normalized to the moment of shock formation defined by
the instant when the entropy behind the shock first exceeds
a value of 3 $k_B$ per nucleon.}
\end{figure*}

\begin{figure*}[!]
\epsfig{file=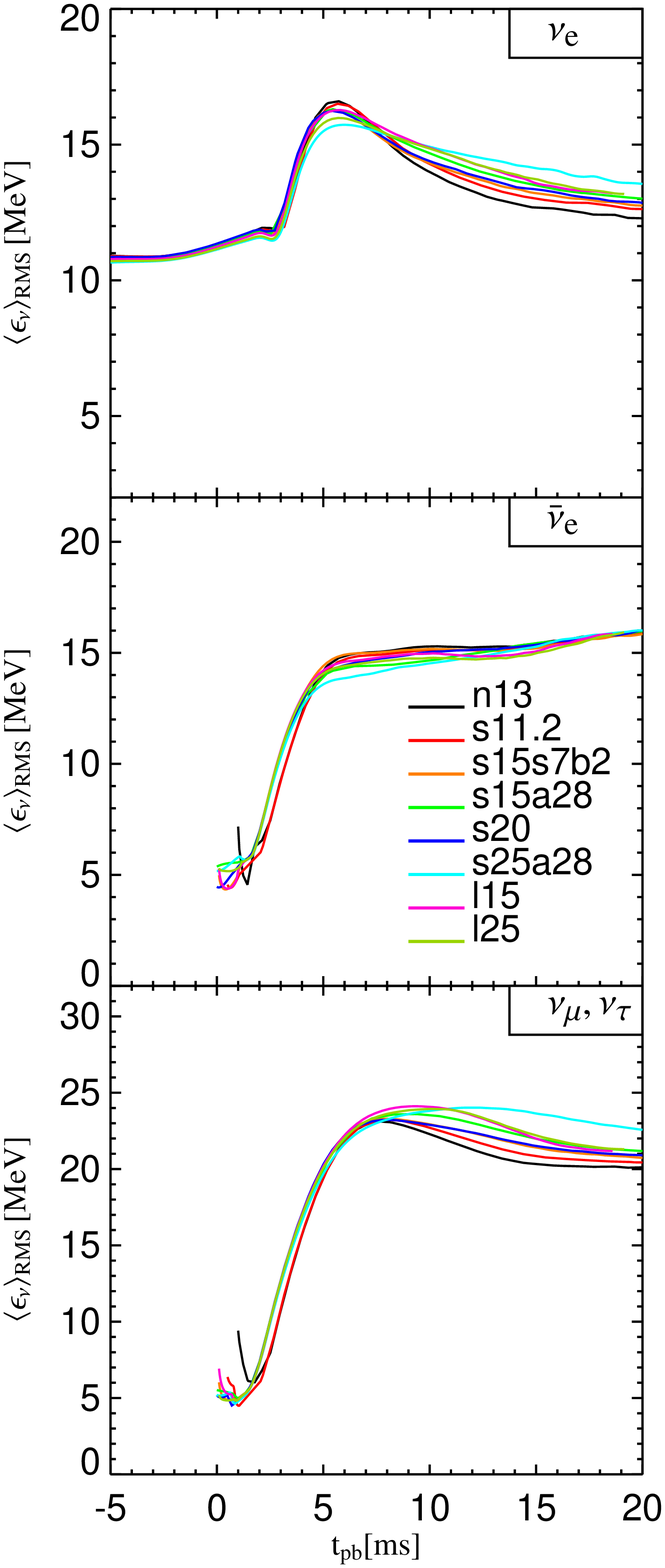,width=0.32\hsize,angle=0}
\epsfig{file=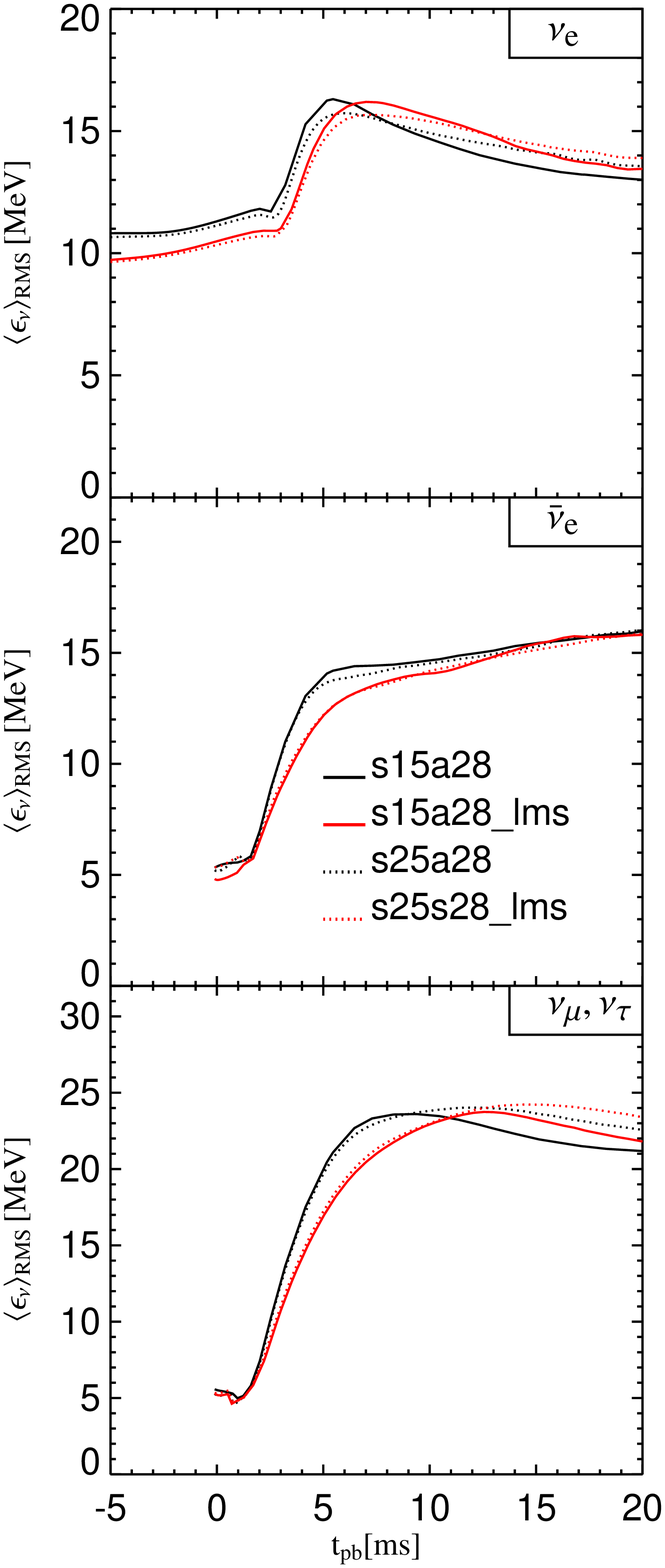,width=0.32\hsize,angle=0}
\epsfig{file=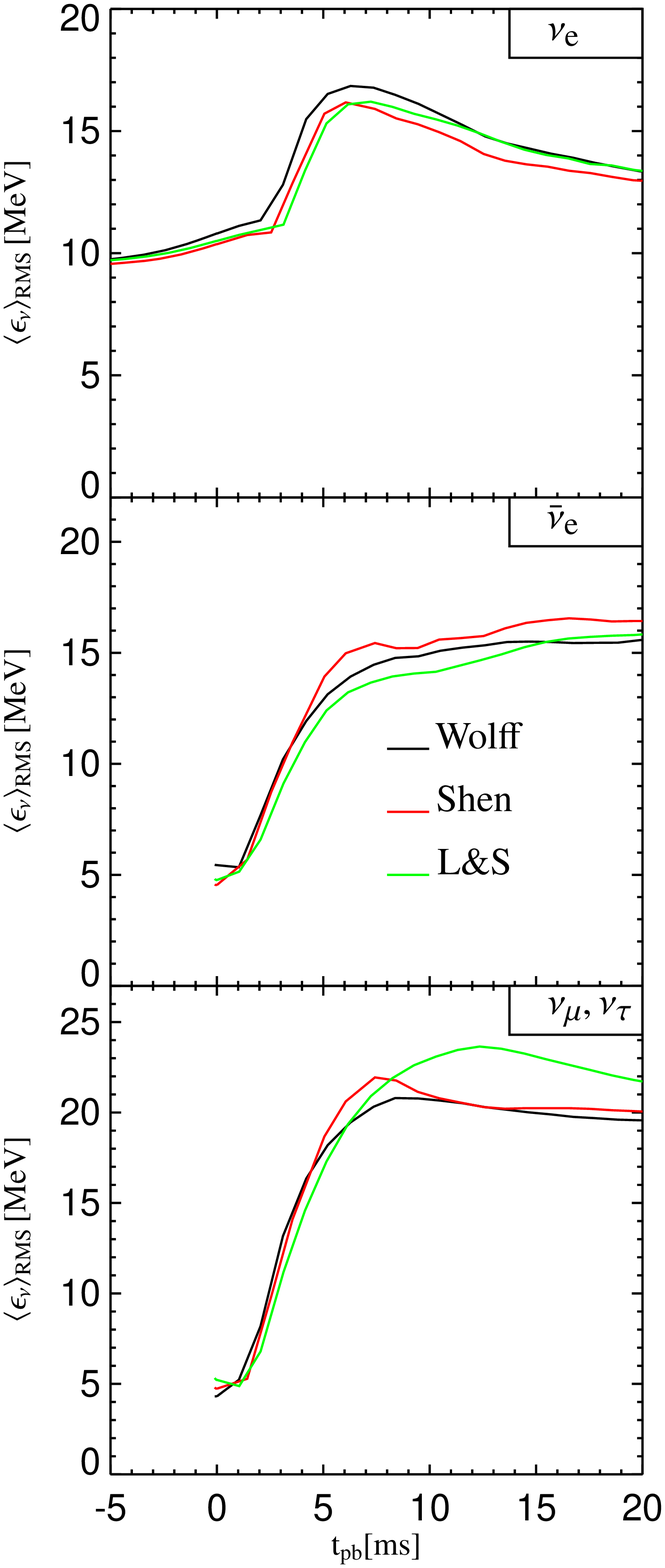,width=0.32\hsize,angle=0}
\caption{\label{erms}
Same as Fig.~\ref{lumi} but for the rms neutrino energies
as defined in Eq.~(\ref{eq:erms}). Note that during core collapse
and in particular before shock formation
only $\nu_e$ were taken into account in the simulations because
the production of $\bar\nu_e$ and heavy-lepton neutrinos is
suppressed due to the low entropy and correspondingly high 
electron and $\nu_e$ degeneracy.}
\end{figure*}

Modern supernova models with sufficiently detailed treatment
of the neutrino physics have in common
the existence of a ``prompt burst'' of electron
neutrinos~\cite{promptburst,Thompson:2002mw,Liebendoerfer:2002ny}. 
This breakout pulse is launched at the moment when the newly formed
supernova shock that races down the density gradient in the
collapsing stellar core reaches densities low enough for
the initially trapped neutrinos to begin streaming faster than
the shock propagates~\cite{Bethe:1980gq}. 
In the shock-heated matter, which is still rich of electrons and
completely disintegrated into free neutrons and protons, 
a large number of $\nu_e$ are rapidly produced by electron
captures on protons. They follow the shock on its way out until
they are released in a very luminous flash, the breakout burst,
at about the moment when the shock penetrates the ``neutrinosphere''
and the neutrinos can escape essentially unhindered.
As a consequence, the lepton number in the layer around the
neutrinosphere decreases strongly and the matter
neutronizes~\cite{burrows_mazurek}. 
Because of the high temperatures
behind the shock, electron-positron annihilation,
nucleon-nucleon bremsstrahlung~\cite{Thompson:2002mw},
and, when $\bar\nu_e$ become more abundant, also neutrino-pair
conversion $\nu_e\bar\nu_e \longrightarrow \nu_{\mu,\tau}\bar
\nu_{\mu,\tau}$~\cite{Buras2003}
are efficient in creating muon and tau neutrino-antineutrino
pairs~\footnote{In the supernova simulations discussed here muon
and tau neutrinos and antineutrinos are treated equally because
their interaction with supernova matter is very similar. This has 
several reasons. On the one hand muons are not present initially 
and at subnuclear densities
in the supernova core and taus cannot be produced at supernova 
conditions. On the other hand neutrino oscillations are suppressed
at the high densities of the core.}. 
The luminosities of the latter therefore begin rising steeply
immediately after shock formation. In contrast, the luminosity
of $\bar\nu_e$ increases more slowly. On the one hand this is
due to the fact that the abundance of positrons and therefore 
the $\bar\nu_e$ production by $e^+$ captures is rather low
as long as electrons are still highly degenerate, on the other
hand pair creation of $\nu_e$ and $\bar\nu_e$ is also suppressed
by the high abundance of $\nu_e$ during the burst phase and the
corresponding fermion blocking in the $\nu_e$ phase space.

These facts can be verified from Figs.~\ref{lumi} and \ref{erms}. 
The rms energies shown in the latter figure are defined by
\begin{equation}
\left\langle\epsilon_{\nu}\right\rangle_{\mathrm{RMS}}\,\equiv\,
\sqrt{\int_0^\infty{\mathrm{d}}\epsilon\int_{-1}^{+1}
{\mathrm{d}}\mu\,f_{\nu}(\epsilon,\mu)\epsilon^5 \over
\int_0^\infty{\mathrm{d}}\epsilon\int_{-1}^{+1}  
{\mathrm{d}}\mu\,f_{\nu}(\epsilon,\mu)\epsilon^3 } \ ,
\label{eq:erms}
\end{equation}
with $f_{\nu}(\epsilon,\mu)$ being the neutrino phase space
distribution, which is a function of the neutrino energy $\epsilon$
and the cosine, $\mu$, of the angle of neutrino propagation relative
to the radial direction.
The results presented in the plots were obtained by core collapse
simulations in spherical symmetry 
with the neutrino-hydrodynamics code developed by Rampp and
Janka~\cite{Rampp:2002bq}, employing a solver for the 
energy-dependent moments equations of neutrino number, energy,
and momentum 
and an approximative treatment of general relativity 
that yields good agreement with fully relativistic simulations,
in particular during the collapse and early postbounce
phases~\cite{Liebendoerfer:2003es}. 

A local minimum in the $\nu_e$ luminosities occurs shortly after
the formation of the shock at core bounce ($t = 0$) and before
the neutronization burst. It is caused by the shock first 
compressing matter from a semi-transparent state to 
neutrino-opaque conditions before the postshock layer
reexpands to become neutrino transparent and to release
the neutronization neutrinos~\cite{Liebendoerfer:2002ny}.
Performing simulations
for a variety of progenitor stars between 11.2$\,M_{\odot}$
and 25$\,M_{\odot}$ from different stellar evolution
modelers~\cite{progenitors}, we have confirmed the uniformity 
of the radiated neutrino luminosities and rms energies in the 
first 20$\,$ms after bounce (Figs.~\ref{lumi} and \ref{erms}, 
left panels) that was also seen
in other recent simulations with neutrino
transport being described by a solution of the
Boltzmann equation or its moments
equations~\cite{Thompson:2002mw,Liebendoerfer:2002ef}.  
The prompt neutronization burst has a typical
full width half maximum of 5--7$\,$ms and a peak luminosity
of 3.3--3.5$\times 10^{53}\,$erg$\,$s$^{-1}$. The striking
similarity of the neutrino emission characteristics 
despite of some variability in the properties of the
pre-collapse cores is caused by a 
regulation mechanism between electron number fraction and
target abundances (protons and nuclei) for electron 
captures~\cite{Bruenn:1985en,messerthesis},
which establishes similar electron fractions in the inner core
during collapse. This leads to a convergence of the structure
of the central part (of roughly a solar mass) of the 
collapsing cores and only small differences in the evolution of
different progenitors until shock breakout~\cite{Liebendoerfer:2002ef}.
Differences of the core size and of the density profile in the
outer part of the iron core and beyond lead to different mass
infall rates at late times when the shock has reached 
neutrino-transparent layers. This implies different accretion
luminosities and thus causes a progenitor-dependent strong
variation of the neutrino emission characteristics after the
$\nu_e$ luminosity has levelled off from the prompt burst.

Only recently improvements in the treatment of electron
capture rates on nuclei during the late phases of stellar
evolution and core collapse have become available which 
remove shortcomings of the widely used independent particle
model in which electron captures 
are suppressed by Pauli blocking for nuclei with
$N\ge 40$~\cite{Bruenn:1985en}.
This typically happens at a density of some
$10^{10}\,$g$\,$cm$^{-3}$ above which electron captures on
free protons govern the evolution of the electron fraction.
The improved rates for core collapse are based on 
Shell Model Monte Carlo calculations of nuclear properties at
finite temperatures, complemented with a Random Phase
Approximation for the 
electron capture rates of a wide sample of nuclei
in the mass range between $A = 65$ and $A = 112$ with 
abundances given by nuclear statistical equilibrium~\cite{Langanke:2003ii}.
In supernova simulations 
with these improved rates electron captures by nuclei
dominate over capture on free protons, and interesting
changes were found during core collapse, bounce, and
postbounce evolution~\cite{Langanke:2003ii,Hix:2003fg}. 
In the panels of the middle columns of Figs.~\ref{lumi} and
\ref{erms} one can see the corresponding differences in 
the neutrino emission properties for simulations 
of a 15$\,M_{\odot}$
and a 25$\,M_{\odot}$ progenitor with the new capture rates
according to Langanke, Mart\'{\i}nez-Pinedo and Sampaio
(LMS)~\cite{Langanke:2001td} 
in comparison to runs with the traditional rate treatment.   
Despite of the visible variations with the rate treatment,
however, the spread of
results for different progenitors does not widen and again
the core properties seem to converge during collapse by a 
self-regulation of electron captures. It is unlikely that
this result will change when incoherent neutrino scattering
off nuclei is included in the models. The effects of this 
process during stellar core collapse have not been satisfactorily
explored yet.

There is still considerable uncertainty in the supernova 
simulations due to our incomplete knowledge of the
nuclear equation of state.
The runs for the different progenitors as well as the 
studies with varied electron capture rates were all
performed with the nuclear equation of state (EoS) of
Lattimer and Swesty (L\&S)~\cite{lsmodel}, 
which is most widely
used in core collapse simulations. It is based on a compressible 
liquid drop model and employs a Skyrme force for the
nucleon interaction. Our choice of the
compressibility modulus of bulk nuclear matter was 180$\,$MeV,   
and the symmetry energy parameter 29.3$\,$MeV, but the 
differences in the supernova evolution caused by other
values of the compressibility of this EoS were shown to be
minor~\cite{Thompson:2002mw,swestymira}.

We have recently tested the effects of the nuclear EoS
by using two available alternative
descriptions~\cite{marekdiploma,Janka:2004tt}, 
on the one hand a new
relativistic mean field EoS (``Shen'')~\cite{shenmodel}
with a compressibility of nuclear matter of 281$\,$MeV and a   
symmetry energy of 36.9$\,$MeV, and on the other hand 
an EoS which was constructed by Hartree-Fock calculations
with a Skyrme force for the nucleon-nucleon interaction
(``Wolff'')~\cite{wolffmodel}
and has a
compressibility of 263$\,$MeV and a symmetry energy of
32.9$\,$MeV. In particular the latter EoS allows for a
faster deleptonization of the less compact and less opaque
postshock layer, thus producing a clearly higher
$\nu_e$ burst in comparison to the standard case with
L\&S EoS. Note that all three runs were performed with
the 15$\,M_{\odot}$ progenitor model s15a28 of Woosley
et al.~\cite{progenitors} and included the use
of the new LMS rates for electron captures. 

Uncertainties of core collapse simulations due to the
use of different numerical schemes for hydrodynamics and
neutrino transport were recently investigated in a 
comparison of the Oak Ridge-Basel and Garching supernova 
codes~\cite{Liebendoerfer:2003es}. Despite of differences
in details, very satisfactory agreement was found for the
overall evolution and for the neutrino emission properties.
The approximative description of general relativity in
the Garching code produces only insignificant deviations
from the fully relativistic treatment of the 
Oak Ridge-Basel code during the infall phase and early 
postbounce evolution including the prompt neutronization 
burst. The burst heights and widths agree nicely between
both codes in Newtonian as well as relativistic runs.

Among the remaining systematic uncertainties in core
collapse models with possible consequences for the
neutronization burst is the unsettled question of
rotation in the progenitor core. While the
large asymmetries seen in supernova explosions
are sometimes claimed to be caused by rapid core rotation
(e.g., Ref.~\cite{Burrows:2004va}),
recent stellar evolution
models seem to disfavor this possibility because they
predict that massive stars lose angular momentum very
efficiently during their evolution. The stellar cores 
therefore rotate so slowly --- rotation periods at the 
edge of the iron core before the onset of gravitational 
instability were determined to be around 100$\,$s ---
that core collapse and bounce remain essentially
unaffected by rotational effects~\cite{Heger:2004qp}.
Instead of rotation, low-mode convective instabilities
have been discussed in this case as potential origin
of the observed global anisotropies of
supernovae~\cite{Blondin:2002sm,Scheck:2003rw}.
For a more reliable determination of the conditions in the
pre-collapse cores, however, truely multi-dimensional
stellar evolution models are desired instead of the currently
used spherical ones that are supplemented by evolution
equations for the lateral averages of the angular momentum
and magnetic field.

If rapid rotation of the pre-collapse iron is still
considered as a viable possibility, despite of probably
valid objections based on current stellar evolution 
models, one may wonder about the effects of such
rotation on the prompt $\nu_e$ burst. For having a noticeable 
influence, the central spin period before collapse must be significantly 
less than 3--5$\,$s, which decreases during collapse
by a factor of about 100. Unfortunately, all numerical
studies of rotational core collapse published so far 
were done with very simplistic or no treatment of neutrino
transport (see, e.g., Ref.~\cite{kotake} and 
references therein), and only the paper by Fryer and
Heger~\cite{Fryer:1999he} provides information in some
detail about the neutrino emission, although the grey
diffusion scheme used in that work falls much behind
the quality and refinement of the transport discussed 
here for simulations of nonrotating (or sufficiently 
slowly rotating) collapsing stars. Besides a possible
dependence of the neutrino signal from the viewing
angle (as a consequence of the rotational deformation
of the core and differences of the shock propagation
and breakout between pole and equator~\cite{moenchmeyer}), the magnitude of 
the neutronization burst and the mean energies of 
neutrinos emitted during the burst seem to be reduced
by rapid rotation~\cite{Fryer:1999he}. For conclusive
results, however, one has to await simulations with a
better treatment of neutrino transport.

\section{Analysis of the neutrino signal}

\subsection{Neutrino fluxes}

The neutrino flux spectra $F_{\nu_i}$ arriving at the Earth are 
determined by the primary neutrino fluxes $F^0_{\nu_i}$ as well as
the neutrino mixing scenario, 
\begin{eqnarray}
\label{eqfluxes1}
F_{\nu_e} & = & p F_{\nu_e}^0 + (1-p) F_{\nu_x}^0 \,, \\ 
F_{\bar\nu_e} & =  &\bar{p} F_{\bar\nu_e}^0 + (1-\bar{p}) F_{\nu_x}^0 ~, \\
4 F_{\nu_x} & = & (1-p) F_{\nu_e}^0 + (1-\bar{p}) F_{\bar\nu_e}^0 +
(2 + p + \bar{p}) F_{\nu_x}^0 \,, \quad
\label{eqfluxes3}
\end{eqnarray}
where $p$ $(\bar p)$ is the survival  probability of an 
electron (anti-)neutrino after propagation through the SN mantle and the
interstellar medium. We restrict our analysis to the standard case of
three active neutrino flavors and negligible magnetic moments or
decays~\footnote{Neutrino decays and magnetic moments 
introduce transitions $\nu_e\to \bar\nu_e$ and are
therefore more easily detectable. For a discussion of these
possibilities see Refs.~\cite{Ando:2003is,Akhmedov:2003fu,Ando:2004qe}.}. 
We assume also that the neutrinos do not cross the Earth before reaching the
detector. The main consequence of Earth matter effects on $\nu_e$
neutrinos---the key channel to observe the neutronization burst---is a
regeneration effect in scenario B and C, thereby slightly improving the
chances to detect the $\nu_e$ burst, while Earth matter effects have
no impact on the signal in scenario A. Therefore, Earth matter effects
increase the differences between scenario A and B/C and it is
conservative to neglect them in our analysis.

The probabilities $p$ and $\bar p$ are basically determined by 
the flavor conversions that take place in the resonance layers, where 
$\rho_{\rm res} \approx m_{\rm N} \Delta m^2_{i}\cos 2\theta / 
(2 \sqrt{2} G_{\rm F} Y_{\rm e} E)$. Here $\Delta m^2_{i}$ and
$\theta$ are the relevant mass difference and mixing angle of the
neutrinos, $m_{\rm N}$ is the nucleon mass,
$G_{\rm F}$ the Fermi constant and $Y_{\rm e}$ the electron fraction.
In contrast to the solar case, SN neutrinos must pass through
two resonance layers: the H-resonance layer at 
$\rho_{\rm H}\sim 10^3$~g/cm$^3$ corresponds to $\Delta m^2_{\rm atm}$,
whereas the L-resonance layer at 
$\rho_{\rm L}\sim 10$~g/cm$^3$ corresponds to $\Delta m^2_{\odot}$
\footnote{The values of the neutrino mass differences used in the
  numerical analysis are $\Delta m^2_{\rm atm}=2.6\times 10^{-3}~{\rm
    eV}^2$ and $\Delta m^2_{\odot}=8.3\times 10^{-5}~{\rm
    eV}^2$.}.
This hierarchy of the resonance densities, along with their
relatively small widths, allows the transitions in the two 
resonance layers to be considered independently~\cite{Dighe:1999bi}.

The neutrino survival probabilities can be characterized by the degree
of adiabaticity of the resonances traversed, which are directly 
connected to the neutrino mixing scheme. 
In particular, the L-resonance is always adiabatic and 
appears only in the neutrino channel, whereas the adiabaticity of
the H-resonance depends on the value of $\theta_{13}$, and 
the resonance appears in the neutrino or anti-neutrino channel 
for a normal or inverted mass hierarchy, respectively.
Table~\ref{tab-pbar} shows the survival probabilities for electron neutrinos,
$p$, and anti-neutrinos, $\bar{p}$, in various mixing scenarios for the
static density profile of the progenitor. Using this profile 
is appropriate, because we are only interested in the
neutrino propagation during the first milliseconds after core-bounce
when the shock wave has not reached the H-resonance yet~\cite{Tomas:2004gr}.
For intermediate values of $\theta_{13}$, i.e.
$10^{-5}\lsim\sin^2 \theta_{13} \lsim 10^{-3}$,
the survival probabilities are no longer constant but depend on the
neutrino energy as well as on the details of the density profile of
the SN.  

\begin{table}
\begin{center}
\begin{tabular}{llccc}
\hline
Scenario & 
Hierarchy &  $\sin^2 \theta_{13}$  &  $p$ &  $\bar{p}$ \\
\hline
A & Normal & $\gsim 10^{-3}$  & 0  & $\cos^2\theta_\odot$ \\
B & Inverted &  $\gsim 10^{-3}$ &  $\sin^2\theta_\odot$ &  0 \\
C & Any & $\lsim 10^{-5}$  & $\sin^2\theta_\odot$ 
&  $\cos^2\theta_\odot$ \\
\hline
\end{tabular}
\caption{
Survival probabilities for neutrinos, $p$, and anti-neutrinos,
$\bar{p}$, for various mixing scenarios in case of the density profile
of the SN progenitor. Terms of the order $\theta_{13}^2$ and 
smaller have been neglected. The solar mixing angle is assumed to be
$\sin^2\theta_\odot=0.30$; for a recent discussion of allowed neutrino
oscillation parameters see Ref.~\cite{Maltoni:2004ei}.
\label{tab-pbar}}
\end{center} 
\end{table} 

For large values of $\theta_{13}$, $\sin^2\theta_{13}\gsim 10^{-3}$, the
H-resonance is adiabatic. In the case of a normal mass hierarchy,
scenario A, the resonance takes place in the neutrino channel. Then
practically all $\nu_e$ which are initially created as $\nu_3$
leave the SN also as $\nu_3$. When they reach the detector, they
have only a small $\nu_e$ admixture, 
$\langle\nu_3|\nu_e\rangle=\sin\theta_{13}$. 
Taking into account the experimental constraints on $\theta_{13}$,
$\sin^2\theta_{13} \lsim 0.047$ at $3\sigma$
C.L.~\cite{Maltoni:2004ei}, one obtains 
$p^{\rm A}=\sin^2\theta_{13}\lsim 0.047$.  
This corresponds to an almost complete
interchange of the $\nu_e$ and $\nu_x$ spectra. 
For an inverted mass hierarchy, scenario B, the resonance occurs 
in the anti-neutrino channel, thus interchanging now almost completely
the $\bar\nu_e$ and $\bar\nu_x$ spectra. In contrast,  the H-resonance is
strongly non-adiabatic for small values of $\theta_{13}$,
$\sin^2\theta_{13}\lsim 10^{-5}$, and for any mass hierarchy (scenario C),
and hence it is ineffective.   
In both the scenarios B and C, the primary $\nu_e$ leave the star
as $\nu_2$ with a large $\nu_e$ admixture at the detector,
$\langle \nu_2|\nu_e\rangle=\sin\theta_\odot$, leading to
$p^{\rm B,C}=\sin^2\theta_\odot=0.30$~\cite{Maltoni:2004ei}. 

Let us assume for simplicity that during the neutronization bursts only
$\nu_e$ neutrinos are emitted. Then the neutrino fluxes arriving
at the Earth are in scenario A
\begin{eqnarray}
 F^{\rm A}_{\nu_e} & = & 
 p^{\rm A} F_{\nu_e}^0 + (1-p^{\rm A}) F_{\nu_x}^0 \approx 0\,, \\ 
 2 F^{\rm A}_{\nu_x} & = & 
 (1-p^{\rm A}) F_{\nu_e}^0 + (1 + p^{\rm A}) F_{\nu_x}^0 \approx
 F_{\nu_e}^0\,,\\
\end{eqnarray}
and in scenario B or C, 
\begin{eqnarray}
 F^{\rm B,C}_{\nu_e} & = & 
 p^{\rm B,C} F_{\nu_e}^0 + (1-p^{\rm B,C}) F_{\nu_x}^0
 \approx \sin^2\theta_\odot F_{\nu_e}^0 \,, \quad\\ 
 2 F^{\rm B,C}_{\nu_x} & = & 
 (1-p^{\rm B,C}) F_{\nu_e}^0 + (1 + p^{\rm B,C})
 F_{\nu_x}^0 \nonumber\\
 & \approx & \cos^2\theta_\odot F_{\nu_e}^0 \,. \quad 
\end{eqnarray}
Hence a detector able to identify
$\nu_e$ events will observe a peak in the $\nu_e$ channel in the cases
B and C, while the peak would be completely absent in case A. 
The signature is similar, although less dramatic, for a detector
observing elastic scattering events. In this case not only $\nu_e$ but
also $\nu_x$ contribute to the signal through neutral-current reactions.
But since the cross section for elastic scattering on
electrons is larger 
for $\nu_e$ than for $\nu_x$ neutrinos, the event number during the
neutronization burst even for such a non-ideal detector is much
larger in the scenarios B and C than in A.

Finally, we note from Fig.~\ref{lumi} that the emission of other 
flavors than $\nu_e$ becomes important already during the end of
the neutronization burst, washing out the big differences expected in
the naive picture above. In the next subsection, we discuss in detail
how the neutronization burst can be identified.

\subsection{Detection of the neutronization burst}

Theoretically, the identification of the neutronization burst is
cleanest with a detector using the charged-current
absorption of $\nu_e$ neutrinos. Examples of such detectors are
heavy water detectors like SNO~\cite{sno} using $\nu_e +d\to e^-
+p+p$, or liquid argon detectors like ICARUS~\cite{Gil-Botella:2003sz}
using  $\nu_e + {}^{40}{\rm Ar} \to e^- + {}^{40}{\rm K}^\ast$.

The simplest possible observable to identify the neutronization burst
is the total number $N$ of $\nu_e$ events within an arbitrary fixed period
$t_{\max}$ after the onset of the neutrino signal.  For instance,
Ref.~\cite{Gil-Botella:2003sz} calculated the expected number of
$\nu_e + {}^{40}{\rm Ar} \to e^- + {}^{40}{\rm K}^\ast$ events in a 70-kton 
liquid argon detector for $t_{\max}=240$~ms, where $t=200$ ms
corresponds to the core bounce, assuming one specific
astrophysical model and as SN distance $d=10$~kpc. They found $N=41$
events in scenario A, compared to $N=86$ in scenarios B and C. 
From the discussion in Sec.~\ref{astro}, it is clear that the
uncertainty in $N$ coming from SN models is rather small. We will
quantify this uncertainty later in Sec.~\ref{models} and use here 10\%
as estimate for the systematic uncertainties due to the SN models. 
Combining these systematic uncertainties and the statistical
fluctuations in quadrature leads to $N=41\pm 8$ for scenario A and
$N=86\pm 13$ for B and C. Thus one could conclude that a
differentiation between the scenarios A and B/C on the 2~$\sigma$
level is possible with a liquid argon detector using the total number
of events. However, we have still neglected another important source
of uncertainty for $N$: The distance to stars in our Galaxy is
typically known only with 25\% accuracy~\cite{distanceaccuracy}. 
The measurement of the SN lightcurve will allow a
determination of its distance with an error of $\approx
5$--10\%~\cite{pm}.  However, the probability that the SN is obscured
by dust is as high as $\sim$75\%. Without an estimate for the SN distance,
the total number of events observed cannot be connected to the SN
luminosity and is thus not a useful observable.   
Instead, we exploit in the following the time structure of the
detected neutrino signal as signature for the neutronization burst.

\begin{center}
\begin{figure}
\epsfig{file=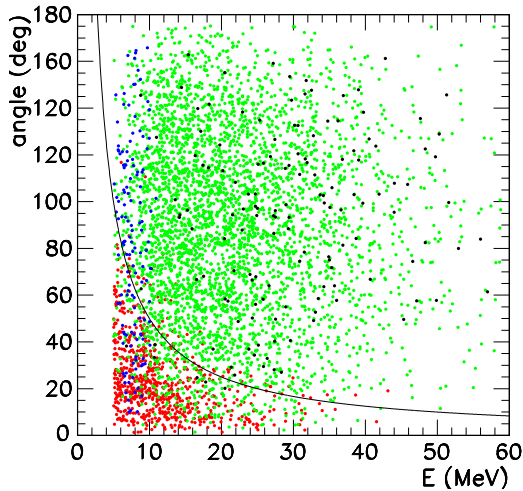,height=7.cm,width=7.cm,angle=0}
\caption{\label{enangle}
Energy and angular distribution of all events with $t<t_{\max}=18$~ms,
assuming $t_{\rm bounce}=0$ ms at bounce,
in a megaton water Cherenkov detector for a SN with a progenitor mass
of $15~M_\odot$ at 5~kpc, and scenario C. The different reaction
channels shown are elastic scattering on electrons (red), 
inverse beta decay (green), 
CC events on oxygen (black) and 
NC events on oxygen (blue); 
also shown is the cut $E({\rm MeV})\times {\rm angle}({\rm degrees})
= 500$.
}
\end{figure}
\end{center}

Since the $\nu_e$ burst lasts only about 25~ms, cf. Fig.~\ref{lumi}, the 
event number in current and proposed charged-current detectors is not high
enough to allow for a detailed time analysis.  Therefore, we will
concentrate in the following on the case of a megaton water Cherenkov
detector, proposed e.g. to be build in Japan~\cite{HK} or in
  the United States~\cite{uno}. 
A drawback of this choice is that this detector type does 
not have a clean signature for the $\nu_e$ channel. 
Instead, one has to consider the $\nu_e$ elastic scattering on
electrons, $\nu_e+e^-\rightarrow \nu_e+e^-$. 
This reaction is basically affected by
three different kinds of backgrounds: inverse beta decay
reactions $\bar\nu_e+p\rightarrow n+e^+$, reactions on 
oxygen, and the elastic scattering of other neutrino flavors on electrons.
In Fig.~\ref{enangle}, we show the distribution of the reconstructed
energies and directions with respect to the vector SN-Earth made of 
all events in a water Cherenkov detector for $t<t_{\max}=18$~ms,
where $t_{\rm bounce}=0$ ms. The events are simulated following
Ref.~\cite{Tomas:2003xn}.

The dominant source of background events are inverse beta
decay reactions. These events are almost isotropically distributed,
while their energy distribution reflects the neutrino energy
spectrum. In contrast, elastic scattering events are concentrated in
the forward direction and at rather low energies. Therefore a cut with
$E({\rm MeV})\times {\rm angle(deg)}<500$ as
shown in Fig.~\ref{enangle} by a solid black line substantially reduces
the number of background events while keeping most of elastic
scattering events. This background is further reduced by using 
in addition Gadolinium to tag the neutrons
produced in the inverse beta reaction~\cite{Beacom:2003nk}.
Reactions on oxygen have a large reaction threshold, $E_{\rm th}>15$~MeV, 
and are therefore not numerous. Moreover, the charged-current events
on oxygen (black dots) have an angular distribution peaked slightly
backwards, and thus the chosen cut eliminates these events
efficiently, too.

\begin{center}
\begin{figure}
\epsfig{file=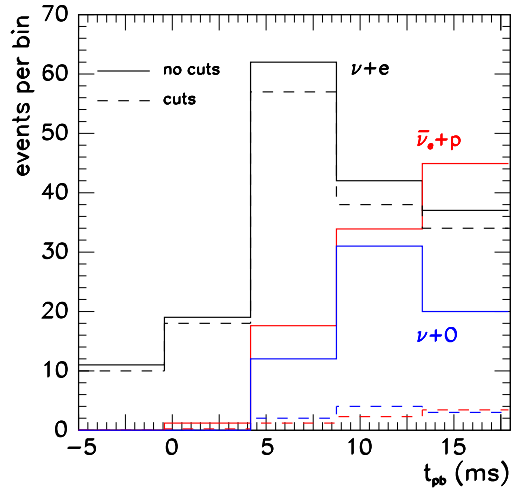,height=7.cm,width=7.cm,angle=0}
\caption{\label{cuts}
Number of events per time bin from the elastic scattering on electrons
(black), inverse beta decay assuming 90\% tagging efficiency
of Gadolinium (red), and reactions on oxygen (blue) in a megaton
water Cherenkov detector for a SN with a progenitor mass of
$15~M_\odot$ at 10~kpc, and case C. Solid lines stand for the number
of events without cut and dashed lines with the cut $(E/{\rm
  MeV})\times ({\rm angle}/{\rm degrees})= 500$.  
}
\end{figure}
\end{center}

As illustration for the efficiency of the background suppression we
show in Fig.~\ref{cuts} the expected signal without (solid lines) and
with cuts (dashed lines). In both cases we assumed an efficiency of
90\% for the Gd tagging of the inverse beta decay reactions.
While the number of elastic scattering events is practically unchanged,
the background of inverse beta decays and reactions on oxygen is
significantly reduced. Therefore, we will consider only the elastic
scattering reactions on electrons in our discussion below. The
sample of elastic scattering events still contains the irreducible
background of scattering on electrons of other neutrinos than
$\nu_e$, but we will show that it is possible to disentangle the
scenarios with and without peak even in the presence of this
background.

\subsection{Results}

In this subsection, we examine in detail how the prompt neutronization
burst from a future galactic SN appears in a water Cherenkov
detector. For this purpose, we have generated elastic scattering events
of neutrinos on electrons using Monte Carlo simulations as described
in Ref.~\cite{Tomas:2003xn}. If not otherwise stated, we have assumed
a megaton detector with energy threshold $E_{\rm th}=5$~MeV and 10~kpc
as the distance to the SN. If the energy threshold could be lowered to
3~MeV, then the event number would typically increase by 20\%. The
neutrino luminosities and energy spectra are based on the SN models
described in Sec.~II.  
In order to follow the time evolution of the signal, we have considered
the time interval from $t=-5$~ms until $t=18$~ms postbounce and divided the
interval into five bins.
As far as the neutrino mixing scenario is concerned, we will compare only
the cases A and C. The first reason for this choice is that the
differences between B and C arising in 
the anti-neutrino channel, $\bar{p}^{\rm B}\neq \bar{p}^{\rm C}$, are always
smaller than their differences to case A. Secondly, 
the mixing scenario B can be confirmed or ruled out by the
modulations in the $\bar\nu_e$ spectrum induced by shock waves in 
the SN envelope or by Earth matter effects, respectively.
However, these modulations are not helpful to distinguish the cases A
and C, since $\bar{p}^{\rm A}= \bar{p}^{\rm C}$ .

\subsubsection{Dependence on the neutrino mixing scenario}

To understand better the results of our Monte Carlo simulations, we
first discuss qualitatively the behavior of 
the expected signal in the presence of the irreducible $\nu_x$
background. The total number of events can be decomposed as
\be
N(t) = N_{\nu_e}(t) + N_{\bar\nu_e}(t) + N_{\nu_x}(t) +
N_{\bar\nu_x}(t) \,,
\ee
where $N_{\nu_i}(t)$ represents the number of events produced in the
reaction $\nu_i+e^-\rightarrow \nu_i+e^-$. Since we are interested in
the differences between scenario A and C, and $\bar{p}^{\rm
  A}=\bar{p}^{\rm C}$,
we will concentrate on the number of events produced by $\nu_e e^-$
and $\nu_x e^-$ scatterings.
Taking into account the linear dependence of the cross sections on the
neutrino energy, $\sigma_{\nu e}(E) \approx \sigma^0_{\nu e} E$, and  
Eqs.~(\ref{eqfluxes1}--\ref{eqfluxes3}), we can write
\begin{eqnarray}
\label{elastic21}
N_{\nu_e}(t) &\propto & \sigma^0_{\nu_e e}\left[p L_{\nu_e}(t)  +
  (1-p) L_{\nu_x}(t)\right] \,, \\
N_{\nu_x}(t) &\propto &  \sigma^0_{\nu_x e}\left[(1-p) L_{\nu_e}(t) +
  (1+p) L_{\nu_x}(t)\right] \,,
\label{elastic22}
\end{eqnarray}
where  $\sigma^0_{\nu_e e}\approx 6\sigma^0_{\nu_x
  e}$. We parameterize our simulation results for 
the different neutrino luminosities by $L_{\nu_x}(t) = \varepsilon(t)
L_{\nu_e}(t)$, where $\varepsilon(t)$ is zero until core bounce 
($t_{\rm pb} = 0$ ms), increases afterwards, and reaches 
$\varepsilon(t) \approx 0.5$ at $t_{\rm pb} = 18$~ms.

\begin{center}
\begin{figure}
\epsfig{file=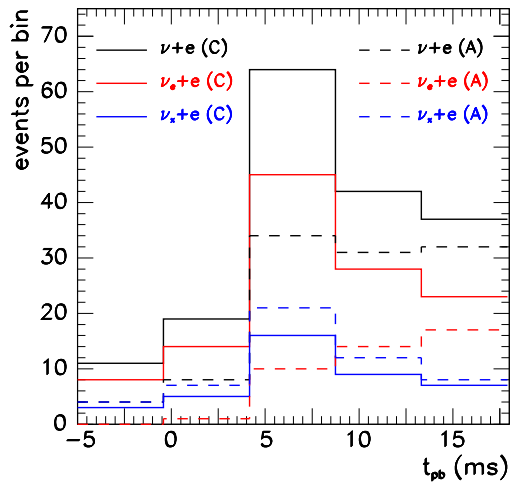,height=7.cm,width=7.cm,angle=0}
\caption{\label{elastic}
Number of events per time bin from the elastic scattering on
electrons: total number (black), $\nu_e~e$ (red) and $\nu_x~e$
(blue), 
for the mixing scenarios C (solid lines) and A (dashed
lines). A megaton water Cherenkov and a SN with progenitor mass 15
$M_\odot$ at 10~kpc is assumed. 
}
\end{figure}
\end{center}

We now examine the difference in the total number of events between scenario
A and C. We define the following ratio, 
\be
 R_\nu^{\rm AC}(t) \equiv 
 \frac{N^{\rm A}_{\nu_e}(t) + N^{\rm A}_{\nu_x}(t)}
      {N^{\rm C}_{\nu_e}(t) + N^{\rm C}_{\nu_x}(t)}
 \approx \frac{1 + 7~\varepsilon(t)}{2.5+5.5~\varepsilon(t)}\,.
\ee
In the first three bins $\varepsilon(t) \approx 0$ and,
therefore, all events are generated from $F^0_{\nu_e}$. In scenario
C, 30\% of the original $\nu_e$ remain as $\nu_e$ whereas 70\%
are converted into $\nu_x$. In case A, 
almost all $\nu_e$ arrive at the Earth as $\nu_x$. Since the cross
section is smaller for $\nu_x e^-$ than for $\nu_e e^-$, fewer events are
expected in scenario A. This difference, though, is not constant
but changes with $\varepsilon(t)$, from $R_\nu^{\mathrm{AC}}(t)\approx 0.4$ 
right before
the bounce until 0.9 at $t_{\rm pb} = 18$ ms. 
This evolution can be clearly observed in
the behavior of the solid (case C) and dashed (case A) black lines in 
Fig.~\ref{elastic}, showing the most important contributions to the
neutronization burst signal, $\nu_e e^-$ (red) and $\nu_x e^-$ (blue), and
the total $\nu e^-$ (black).

Next, we study the different contributions from $\nu_e$ and
$\nu_x$  scatterings to the signal. In particular, from
Eqs.~(\ref{elastic21}--\ref{elastic22}) we define the ratio 
\be
R_{ex}(t) \equiv \frac{N_{\nu_e}(t)}{N_{\nu_x}(t)} \approx
6 \: \frac{p+(1-p)~\varepsilon(t)}{1-p + (1+p)~\varepsilon(t)}\,.
\ee 
In case C, $R^{\rm C}_{ex}(t) > 1$ during the whole burst. Therefore,
$\nu_e e^-$ scatterings generate always more events than $\nu_x e^-$
reactions, as can be seen in Fig.~\ref{elastic}. This is very different in
case A: The value of the ratio, 
$R^{\rm A}_{ex}(t)\approx 6\varepsilon(t)/(1+\varepsilon(t))$, strongly
depends on time. In the first two bins, $\varepsilon(t)\approx
0$, and thus $R^{\rm A}_{ex}(t) \approx 0$. Therefore, practically all events
are generated by $\nu_x e^-$. At $t_{\rm pb}\gsim 11$~ms, however,
$\varepsilon(t)$ has increased so much that $R^{\rm A}_{ex} \gsim 1$. In
Fig.~\ref{elastic}, we
can see how in the first three bins the dashed blue lines are above
the red ones, but in the fourth bin both lines cross and events from
$\nu_e e^-$ scatterings become more important. 
This feature will play a key role in disentangling both
mixing scenarios.

Finally, we discuss the time evolution of the signal. This evolution
results from an interplay of the time dependence of $L_{\nu_e}(t)$ and 
$L_{\nu_x}(t)$. Whereas $L_{\nu_e}(t)$ shows a clear peak structure
around $t_{\rm pb} \approx 7$ ms, $L_{\nu_x}(t)$ steadily increases
after the bounce. 
In order to  discuss which time behavior dominates, we consider
separately the two channels $N_{\nu_e}(t)$ and $N_{\nu_x}(t)$. From
Eqs.~(\ref{elastic21}--\ref{elastic22}), we can estimate the ratio
between events connected to $L_{\nu_x}(t)$ and to $L_{\nu_e}(t)$. 
In scenario C, this ratio is smaller than one for both channels
$N_{\nu_e}(t)$ and $N_{\nu_x}(t)$. Thus the component generated by
$L_{\nu_e}(t)$, and therefore producing a clear peak, dominates over
$L_{\nu_x}(t)$. This is reflected in Fig.~\ref{elastic} in the peak
structure of both the solid red and blue lines. 
In scenario A, the ratio is smaller than one in $N_{\nu_x}(t)$,
but it is larger that one in $N_{\nu_e}(t)$. Therefore, $L_{\nu_e}(t)$
dominates over $L_{\nu_x}(t)$, if $R^{\rm A}_{ex}(t)<1$.
As a consequence, the net result in scenario C is an enhanced peak, whereas 
in scenario A the time structure is more complicated. In the first three
bins, $N_{\nu_x}(t)$ dominates over $N_{\nu_e}(t)$ producing a peak
like in case C, but with fewer events. In the fourth and fifth bin, however,
$N_{\nu_e}(t)$ becomes larger and there is a cancellation between the negative
slope of $N_{\nu_x}(t)$ and the positive of $N_{\nu_e}(t)$. The final
result is an almost flat structure in case A (dashed black lines in 
Fig.~\ref{elastic}), in contrast to the clear decrease of events in
the last two bins in case C.

\begin{center}
\begin{figure}
\epsfig{file=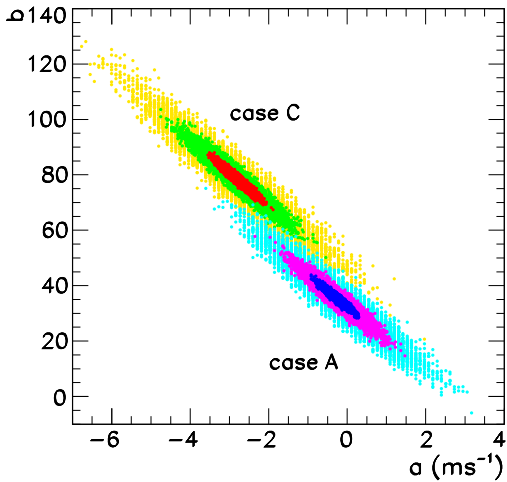,height=7.cm,width=7.cm,angle=0}
\caption{\label{a-b}
Distribution of fit values $a$ and $b$ for different SN
distances, $D=2,5$ and 10~kpc, and a $15~M_\odot$ SN progenitor.} 
\end{figure}
\end{center}

In order to illustrate the difference in the slope after the third bin
we have simulated the neutrino signal from 10000 SN explosions for a
progenitor with $15~M_\odot$. Then we have fitted the
last three bins in each case by a straight line, $y = at/{\rm ms}+b$.  
In Fig.~\ref{a-b}, we show the normalized distribution of the fit
values of $a$ and $b$ for scenarios A and C and three different
distances, $D=2,5$ and 10~kpc. The slope, $a$, is centered at almost 0
ms$^{-1}$ in mixing scenario A, whereas in C the center lies at
about $-3~{\rm ms}^{-1}$.

\subsubsection{Dependence on the progenitor mass and SN models}
\label{models}
We have discussed how the presence or absence of a peak in the neutrino
signal during the neutronization burst is related to the neutrino
mixing scenario. In this subsection, we analyze the robustness
of this connection, studying its dependence on several SN parameters.

\begin{center}
\begin{figure}
\epsfig{file=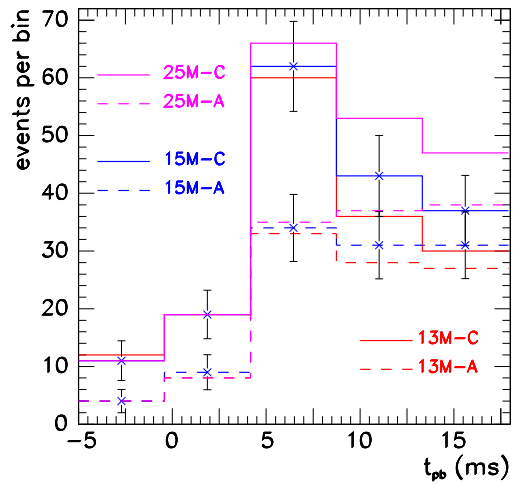,height=7.cm,width=7.cm,angle=0}
\caption{\label{Mdep}
Number of events per time bin for different reactions in a megaton
water Cherenkov detector for a SN at 10 kpc for cases A (dashed lines)
and C (solid lines) and for different progenitor masses: $13~M_\odot$ (n13)
in red, $15~M_\odot$ (s15s7b2)  in magenta, and $25~M_\odot$ (s25a28) in
blue. Statistical errors are also shown for the $15~M_\odot$ case.
}
\end{figure}
\end{center}

If the SN is optically obscured, then the progenitor cannot be
identified. Thus we have to ensure that the neutrino signal during the
$\nu_e$ burst depends only weakly on the progenitor model.
We have analyzed the expected neutrino signal for various progenitors with
main sequence masses $11.2~M_\odot$ (s11.2), $13~M_\odot$ (n13), 
$15~M_\odot$ (s15s7b2), $15~M_\odot$ (s15a28), $20~M_\odot$ (s20) and 
$25~M_\odot$ (s25a28)~\cite{progenitors}.
In Fig.~\ref{Mdep}, we show for illustration the results for two extreme
progenitors, n13 and s25a28, as well as for an intermediate case,
s15s7b2.

We find that for all progenitor masses the peak structure described in
the last subsection can be clearly observed in  case C but not in
case A. In the first three bins the differences between different
progenitor masses are smaller than the statistical fluctuations. 
The average event number per bin varies less than  6\% changing the
progenitor mass from $11~M_\odot$ to $25~M_\odot$. This variation is
at least a factor two smaller than the statistical fluctuations.
However, after the third bin the predictions for different progenitor
models vary more strongly, reaching 10\% or the size of the statistical
fluctuations. At this time, neutrinos other than $\nu_e$ start to play an
important role and the predictions become more model-dependent,
cf. Fig.~\ref{lumi}. 
These differences, though, do not affect the observation of the 
neutronization peak, and therefore do not spoil the possibility to
disentangle the scenarios A and C.

\begin{center}
\begin{figure}
\epsfig{file=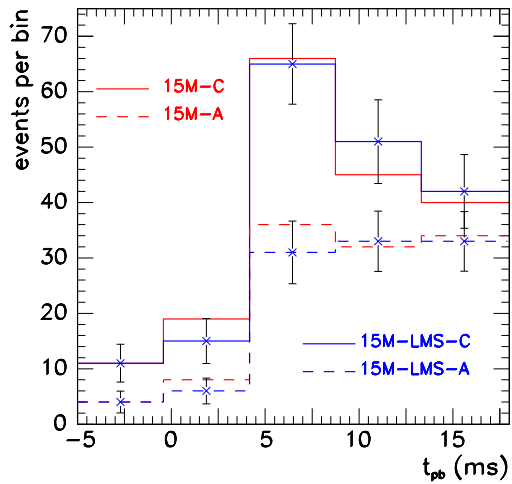,height=7.cm,width=7.cm,angle=0}
\caption{\label{ecapturedep}
Same as Fig.~\ref{Mdep} for mixing scenarios A (dashed lines) and C
(solid lines) for a SN at 10 kpc and a progenitor mass of
$15~M_\odot$ (s15a28). In blue we show the SN model with the electron
capture rates on heavy nuclei adopted from 
Ref.~\cite{Langanke:2003ii} (LMS), and
in red with the traditional description. Statistical errors are also shown for
the (LMS) models.
}
\end{figure}
\end{center}

Another source of uncertainty in our predictions of the neutrino signal
may be the incomplete or inaccurate treatment of some of the
weak interaction rates, in particular with nuclei, used in the SN simulations. 
As an example for the changes in the neutrino signal induced by an
improved treatment of these rates we consider the effect of
electron captures by nuclei in the core.
We compare the signal expected for the progenitor models s15a28 and
s25a28, see Fig.~\ref{ecapturedep}, with and without including 
the electron capture rates on nuclei with mass numbers larger than 65 as
suggested in Ref.~\cite{Hix:2003fg} (LMS).
In both s15a28 and s25a28 SN models we observe the main differences in
the central bins. This change is directly connected to the behavior of the
luminosities: Electron capture on heavy nuclei broadens the peak
in $L_{\nu_e}(t)$ and reduces somewhat $L_{\nu_x}(t)$. However, the changes 
are always smaller than the statistical fluctuations and therefore
do not affect the observation of the neutronization peak.

\begin{center}
\begin{figure}
\epsfig{file=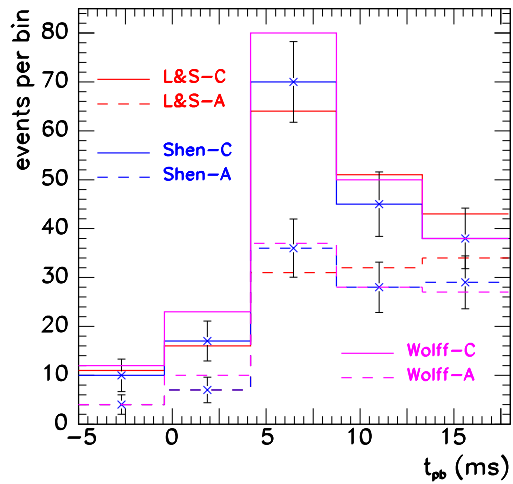,height=7.cm,width=7.cm,angle=0}
\caption{\label{EOSdep}
Same as Fig.~\ref{Mdep} for mixing scenarios A (dashed lines) and C
(solid lines) for a SN at 10 kpc and a progenitor mass
of $15~M_\odot$ (s15a28). 
We show the SN models computed with the equations of
state ``L\&S'' (red), ``Shen'' (blue), and ``Wolff'' (magenta),
respectively. Statistical errors are displayed for the ``Shen'' model.
}
\end{figure}
\end{center}

For the SN progenitor s15a28 we have also investigated the influence
of the nuclear equation of state (EoS) on the evolution of the
luminosity during the neutronization burst. We have considered three
different models for the EoS, dubbed ``L\&S''~\cite{lsmodel}, 
``Shen''~\cite{shenmodel}, and ``Wolff''~\cite{wolffmodel}.
In Fig.~\ref{EOSdep} we show the predicted numbers of events depending
on the employed EoS. The main change is the total luminosity in the
peak. In case C, the change in the luminosities results only in a
rescaling of the total number of events in each bin. In case A,
though, the increase of $L_{\nu_e}$ leads  
to an enhancement of the $\nu_x+e^-$ contribution, whereas the $\nu_e+e^-$
events are not strongly affected. As can be seen in
Fig.~\ref{elastic}, this may slightly modify the time-structure 
and may lead to a small peak in the third bin. However, these changes
are again much smaller than the size of the statistical fluctuations.

\begin{center}
\begin{figure}
\epsfig{file=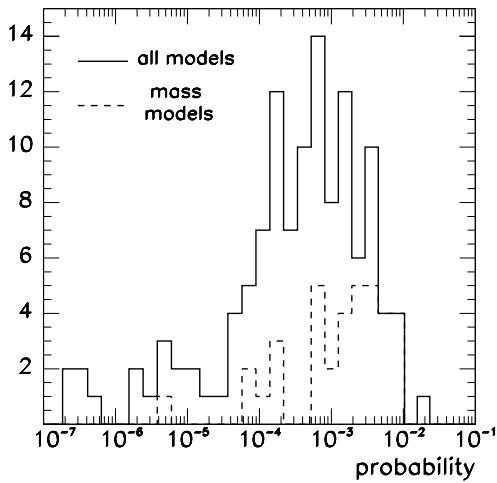,height=7.cm,width=7.cm,angle=0}
\caption{\label{chi2}
Distribution of the probability to misidentify cases A and C for
all SN models considered in the previous sections combined
(solid lines), and only for those with different progenitor mass (dashed
lines). 
}
\end{figure}
\end{center}

In summary, we have found that the changes in the predicted event
numbers for different progenitor models, improvements in the interaction
rates or changes in the EoS are small compared to the size of the
statistical fluctuations. However, up to now we have always varied
only one parameter, e.g. the progenitor mass, and kept all others
fixed. In order to quantify the probability to disentangle the neutrino
mixing scenarios A and C, we should in principle first quantify the
uncertainties in all input parameters of the SN model and then derive
the resulting uncertainties in the neutrino fluxes. Already the first
step is impractical. Therefore, we use the following method: We take
all the SN models considered in the previous sections and calculate
for each the expected number of events for cases A,
$N_{\alpha,i}^{\rm A}$, and C, $N_{\alpha,i}^{\rm C}$, where
$\alpha$ refers to a SN model. Then we compute the probability that
scenario A and C can be distinguished for all possible pairs
$\alpha,\beta$ using a $\chi^2$ analysis, 
\begin{equation}
 \chi^2_{\rm AC}(\alpha,\beta) = \sum_{\rm i=1}^5
 \frac{(N_{\alpha,i}^{\rm A}-N_{\beta,i}^{\rm C})^2}
      {N_{\alpha,i}^{\rm A}+N_{\beta,i}^{\rm C}} \,.
\end{equation}
In Fig.~\ref{chi2} we plotted with solid lines the distribution of this
probability using all possible combinations of the SN models introduced
previously. In the worst combination, scenario A and C can be
distinguished only at the $2\sigma$ level ($p=4.5\%$). However, in
most cases the probability to misidentify scenarios A and C is smaller
than $3\sigma$ ($p=0.27\%$). 
We show in Fig.~\ref{chi2} also the case that the systematic
uncertainty of the SN models is dominated by the unknown progenitor
star (dashed line). Then it is possible to
  disentangle the scenarios A and C with a confidence level
  better than $3\sigma$ for most combinations.

\subsubsection{Intermediate values of $\theta_{13}$ and scenario B}

\begin{center}
\begin{figure}
\epsfig{file=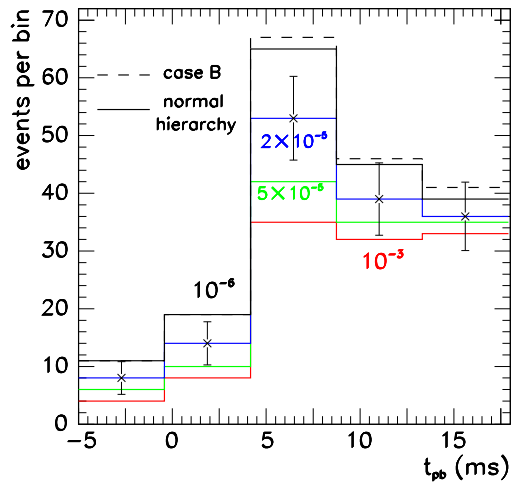,height=7.cm,width=7.cm,angle=0}
\caption{\label{theta13dep}
Same as Fig.~\ref{Mdep} for a SN at 10 kpc and a progenitor mass
of $15~M_\odot$ (s15a28), for different neutrino parameters.
 We show with a dashed line case B, inverted mass hierarchy, and with
 solid lines a normal mass hierarchy for different values of
 $\sin^2\theta_{13}$. Statistical errors are shown for
 $\sin^2\theta_{13}=2\times 10^{-5}$.} 
\end{figure}
\end{center}

For completeness, we analyze in this subsection the dependence of the
neutrino signal on intermediate values of $\theta_{13}$ in scenario A,
and the case of scenario B. For this purpose we
fix a SN progenitor model, s15a28, and consider first mixing scenarios
that lie between A and C, i.e. a normal mass hierarchy and
$\sin^2\theta_{13}=10^{-3}$, $5\times 10^{-5}$, $2\times 10^{-5}$ and
$10^{-6}$, see Fig.~\ref{theta13dep}. The first and the last value
correspond to scenario A and C, respectively.
Whereas $\bar{p}(E)$ is independent on $\theta_{13}$, the average
value of $p(E)$ grows continuously from 0 in mixing scenario A to
$\sin^2\theta_\odot\approx 0.3$ in scenario C.  
This behavior of the $\nu_e$ survival probability results in a
continuous increase of the peak.
For instance, the probability to disentangle the case of 
intermediate $\theta_{13}$ from scenario A becomes smaller than
2$\sigma$ for $\sin^2\theta_{13}\gsim 10^{-5}$ and the SN progenitor
s15a28. Therefore, the detection of a peak not only excludes case A
($\sin^2\theta_{13}\gsim 10^{-3}$), but also most of the intermediate
range of $\theta_{13}$, $\sin^2\theta_{13}\gsim 10^{-5}$.

Finally, we show also an example of case B, inverted mass hierarchy
and large $\theta_{13}$, $\sin^2\theta_{13}=10^{-2}$. 
Since the survival probability for anti-neutrinos is different for B and C, 
the total number of events also differs slightly. 
In particular, the number of events in case B is expected to be
larger, as it can be seen in Fig.~\ref{theta13dep}. The reason is
twofold: First, $L_{\bar\nu_x}>L_{\bar\nu_e}$, and second
$\sigma_{\bar\nu_e e}>\sigma_{\bar\nu_x e}$. Since $\bar{p}^{\rm B}\approx
0$, more $\bar\nu_x$ are converted into $\bar\nu_e$ than in case C, and
 the larger $\bar\nu_e~e^-$ cross section implies more events.
Hence, we conclude that scenario B is slightly easier to distinguish from C
than A from C.

\section{Distance determination}

After the neutrino mixing scenario has been determined, the small
spread in the predicted total number of events $N$ suggests that $N$
is a useful estimator for the SN distance $D$. Since $D\propto N^{1/2}$,
the relative uncertainty $\delta N/N_0$ translates into 
\be
\delta D = \frac{D_0}{2N_0} \delta N \,.
\ee
For each SN model $\alpha$, the probability distribution $p_\alpha(N)$
of the observed number of events $N$ is
Gaussian with $\sigma_{\rm stat}=\sqrt{\langle N_\alpha\rangle}$. To estimate
the ``systematic'' uncertainty $\sigma_{\rm syst}$ due to differences
in the SN models, we use the following recipe: We calculate for different 
models the individual expectation value $\langle N_\alpha\rangle$ and use
then their variance as systematic uncertainty, i.e.
\be
 \sigma_{\rm sys}^2 = 
 \frac{1}{m-1}\sum_{\alpha=1}^m 
 \left( \langle N_\alpha\rangle - \langle N\rangle\right)^2 \,,
\ee 
where $m$ stands for the total number of SN models considered, and 
$\langle N\rangle=(1/m)\sum_\alpha  \langle N_\alpha\rangle$. 

\begin{center}
\begin{figure}[]
\epsfig{file=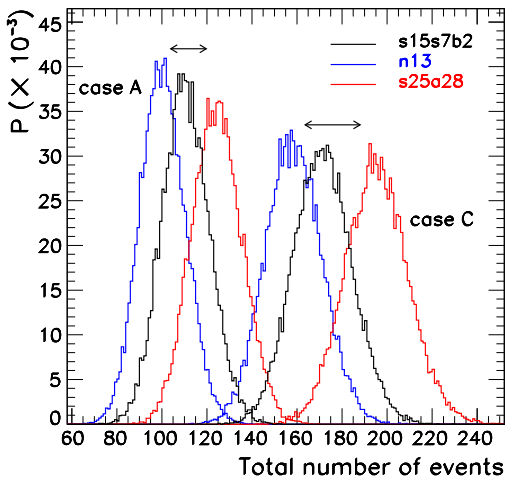,height=7.cm,width=7.cm,angle=0}
\caption{\label{delta_N_mass}
Distribution $p_\alpha(N)$ of the observed total
number of events $N$ for 20000 simulated SNe with
  progenitors}: n13, s15s7b2, and s25a28, and for the neutrino mixing
scenarios A and C. A distance to the SN of 10 kpc has been assumed.
\end{figure}
\end{center}

The two most important sources of systematic errors are the nuclear
equation of state, $\sigma^{\rm EoS}_{\rm sys}$ and, if the SN 
progenitor is not identified, its mass, $\sigma^{\rm mass}_{\rm sys}$.
In order to quantify the influence of different progenitor types we have
considered six different SN models (s11.2, n13, s15s7b2, s15a28, s20,
and s25a28) discussed in Sec.~II.
For the time interval considered here, from $t=-5$~ms until $t=18$~ms
post bounce, we obtain as systematic error $\sigma^{\rm mass}_{\rm
  sys} \approx 7$\%. In the case of a SN located at 10 kpc we expect
as average value $\langle N\rangle=112$ and $\sigma^{\rm mass}_{\rm
  syst}\approx 8$ as systematic error for scenario A, compared to 
$\langle N\rangle=176$ and $\sigma^{\rm mass}_{\rm syst}\approx 12$ for
scenario C. 
In Fig.~\ref{delta_N_mass}, we show the probability distribution
$p_\alpha(N)$ of the observed number of events $N$ until $t=18$~ms
post bounce, choosing as examples three SN models: s15s7b2, and the
extreme cases n13 and  s25a28. 
The obtained range 
\mbox{$[\langle N\rangle-\sigma^{\rm mass}_{\rm sys}: \langle
  N\rangle+\sigma^{\rm mass}_{\rm 
  sys}]$} is also shown by arrows for scenarios A and C.

\begin{center}
\begin{figure}[]
\epsfig{file=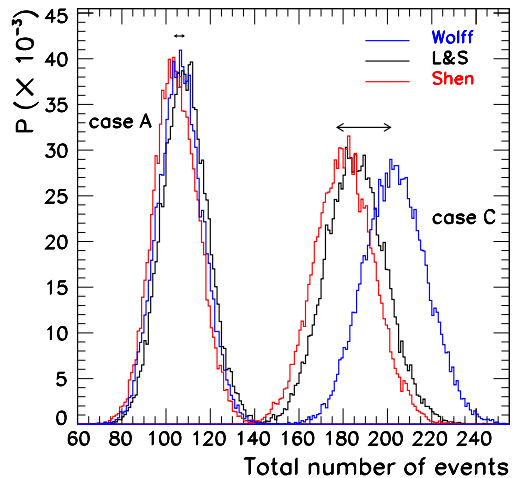,height=7.cm,width=7.cm,angle=0}
\caption{\label{delta_N_eos}
Same as Fig.~\ref{delta_N_mass} for the SN progenitor s15a28 and three
different EoS: L\&S, Shen and Wolff.
}
\end{figure}
\end{center}

The second source of systematic uncertainty analyzed is the nuclear
equation of state. As an illustration of its effect on the
determination of the SN distance we have considered the SN
progenitor s15a28 using the three different EoS defined in Sec.~II:
L\&S, Shen and Wolff.
While the energy emitted in $\nu_e$ and $\nu_x$ until $t=18$~ms
post bounce in the models 
L\&S and Shen differs by less than 3\%, the energy released in
the Wolff model differs by  more than 16\%. This difference depends
on the flavor: in the Wolff model more $\nu_e$'s  but
fewer $\nu_x$'s are emitted than in the other models.
In scenario C the signal is dominated by $L_{\nu_e}$, and thus the
number of events in the Wolff model is much larger than in
the other two models. However, in scenario A the contribution from
$L_{\nu_x}$ is of the same order of magnitude as that of
$L_{\nu_e}$, and since $L_{\nu_x}$ is smaller in 
the Wolff model, this compensates the increase in the number of events
due to $L_{\nu_e}$. This leads to much less variation in the total
number of events in case A than in C.
For the same time interval we have obtained a systematic error
$\sigma^{\rm EoS}_{\rm sys} \approx 2$\% and 6\% in cases A and C,
respectively.  
If a SN is located at 10 kpc we expect as average value
$\langle N\rangle=106$ and $\sigma^{\rm EoS}_{\rm syst}\approx 2$ as systematic
error for scenario A, compared to  
$\langle N\rangle=189$ and $\sigma^{\rm EoS}_{\rm syst}\approx 12$ for
scenario C. 
In Fig.~\ref{delta_N_eos}, we show the probability distribution
$p_\alpha(N)$ of the observed number of events $N$ until $t=18$~ms
post bounce for the three different models considered.
The corresponding range 
\mbox{$[\langle N\rangle-\sigma^{\rm EoS}_{\rm sys}: \langle
  N\rangle+\sigma^{\rm EoS}_{\rm 
  sys}]$} is also shown by arrows for scenario A and C.

Assuming that $\sigma^{\rm mass}_{\rm sys}$ and 
$\sigma^{\rm EoS}_{\rm sys}$ dominate the total systematic uncertainty  
$\sigma_{\rm sys}$ we obtain a total ``systematic error'' of 7\% and
9\% for scenarios A and C, respectively. 
We can now combine the statistical errors associated with the
averages $\langle N \rangle$ and systematical errors again in
quadrature. For the cases discussed in this section this 
results in a relative error of 12\% for the
total event number, or a 6\%
error in the SN distance as predicted from the prompt
neutronization burst. 

This result represents a significant improvement with respect to the
typical accuracy of the distance determination for stars 
in our Galaxy, which is of
the order of 25\%. In the case that the SN lightcurve can be
measured, the distance to the star could be determined with a similar
accuracy, 5--10\%. However, it is likely that the next Galactic 
SN is obscured by 
interstellar dust: about three out of four SNe are estimated to be
optically obscured by dust~\cite{pm}. In this case the measurement of
the neutrino signal from the 
neutronization burst would provide the only way to determine the
distance to the star.

\section{Summary}

We have performed simulations of SN explosions for a variety of
progenitor models and differences in the underlying microphysics used.  
The resulting changes in the neutrino signal from the neutronization
burst were always small compared to the statistical fluctuations
expected for a megaton water Cherenkov detector.
We have argued that the total number of events is not a useful
observable, if the SN is optically obscured by dust. Instead, we
propose to use the time structure of the neutrino signal as discriminator
for the neutrino mixing scenario: the observation of a peak in the
first milliseconds of the neutrino signal would rule out the normal mass
hierarchy with ``large'' $\theta_{13}$ (case A), cf. Figs.~6--9.
Thereby, the degeneracy between scenarios A and C, which both produce
the same $\bar\nu_e$ signal during the accretion and cooling phases,
can be broken.

After the neutrino mixing scenario has been established, the small
uncertainty in the predicted value of the total event rate during the
neutronization burst phase makes a determination of the SN distance
feasible independent from optical observations. Provided that the
predictions of current progenitor models and state-of-the-art
simulations do not miss important physics, we estimate that the SN
distance can be measured with an accuracy of $\sim$6\%.
Since it is likely that a Galactic SN is optically
obscured by interstellar dust and no other distance determination can
be used, this new method relying only on neutrinos seems very
promising.

\begin{acknowledgments}

We would like to thank A.~Dighe and G.~Raffelt for many useful
discussions and K.~Langanke, G.~Mart\'{\i}nez-Pinedo and J.M.~Sampaio
for their table of electron capture rates on nuclei, which was calculated 
by employing a Saha-like NSE code for the abundances written by
W.R.~Hix.
MK acknowledges support by an
Emmy-Noether grant of the Deutsche Forschungsgemeinschaft and RT by a
Marie-Curie-Fellowship of the European Community. Support of the
Sonderforschungsbereich 375 ``Astro Particle Physics'' of the 
Deutsche Forschungsgemeinschaft is acknowledged, too.

\end{acknowledgments}



\begin{thebibliography}{99}

\bibitem{Dighe:1999bi}
A.~S.~Dighe and A.~Y.~Smirnov,
Phys.\ Rev.\ D {\bf 62}, 033007 (2000)
[hep-ph/9907423].


\bibitem{Lunardini:2003eh}
C.~Lunardini and A.~Y.~Smirnov,
JCAP {\bf 0306} (2003) 009
[hep-ph/0302033].

\bibitem{Takahashi:2003bj}
K.~Takahashi and K.~Sato,
Nucl.\ Phys.\ A {\bf 718} (2003) 455.

\bibitem{Raffelt2001}
G.~G.~Raffelt,
Astrophys.\ J.\  {\bf 561}, 890 (2001)
[astro-ph/0105250];

\bibitem{Buras2003}
R.~Buras, H.-T.~Janka, M.~T.~Keil, G.~G.~Raffelt and M.~Rampp,
Astrophys.\ J.\  {\bf 587}, 320 (2003)
[astro-ph/0205006].

\bibitem{Keil2003}
M.~T.~Keil, G.~G.~Raffelt and H.-T.~Janka,
Astrophys.\ J.\  {\bf 590}, 971 (2003)
[astro-ph/0208035].

\bibitem{earth}
C.~Lunardini and A.~Y.~Smirnov,
Nucl.\ Phys.\ B {\bf 616}, 307  (2001)
[hep-ph/0106149];
K.~Takahashi and K.~Sato,
Phys.\ Rev.\ D {\bf 66}, 033006 (2002)
[hep-ph/0110105];
A.~S.~Dighe, M.~T.~Keil and G.~G.~Raffelt,
JCAP {\bf 0306}, 006 (2003)
[hep-ph/0304150];
A.~S.~Dighe, M.~Kachelrie\ss, G.~G.~Raffelt and R.~Tom\`as,
JCAP {\bf 0401}, 004 (2004)
[hep-ph/0311172].

\bibitem{shock}
R.~C.~Schirato and G.~M.~Fuller, 
astro-ph/0205390;
G.~L.~Fogli, E.~Lisi, D.~Montanino and A.~Mirizzi,
Phys.\ Rev.\ D {\bf 68} (2003) 033005
[hep-ph/0304056].

\bibitem{Tomas:2004gr}
R.~Tom\`as, M.~Kachelrie\ss, G.~Raffelt, A.~Dighe, H.-T.~Janka and L.~Scheck,
JCAP {\bf 0409}, 015 (2004)
[astro-ph/0407132].


\bibitem{burst}
For a review, see A.~Burrows, ``Neutrinos from Supernovae'',
in {\it Supernovae}, A.~G.~Petschek (Ed.), Springer, New York (1990), 
p.~143.

\bibitem{Mueller:2003fs}
E.~M\"uller, M.~Rampp, R.~Buras, H.-T.~Janka and D.~H.~Shoemaker,
``Towards gravitational wave signals from realistic core collapse supernova
models,''
Astrophys.\ J.\  {\bf 603}, 221 (2004)
[astro-ph/0309833].


\bibitem{promptburst}
S.~W.~Bruenn,
Phys.\ Rev.\ Lett.\  {\bf 59}, 938 (1987);
R.~Mayle, J.~R.~Wilson and D.~N.~Schramm,
Astrophys.\ J.\  {\bf 318}, 288 (1987);
E.~S.~Myra and A.~Burrows,
Astrophys.\ J.\  {\bf 364}, 222 (1990);
M.~Rampp and H.-T.~Janka,
Astrophys.\ J.\  {\bf 539}, L33 (2000)
[astro-ph/0005438];
M.~Liebend\"orfer, A.~Mezzacappa, F.~K.~Thielemann, O.~E.~B.~Messer,
W.~R.~Hix and S.~W.~Bruenn, 
Phys.\ Rev.\ D {\bf 63}, 103004 (2001)
[astro-ph/0006418];
M.~Liebend\"orfer, O.~E.~B.~Messer, A.~Mezzacappa, S.~W.~Bruenn,
C.~Y.~Cardall and F.~K.~Thielemann,
Astrophys.\ J.\ Suppl.\ {\bf 150}, 263 (2000).

\bibitem{Thompson:2002mw}
T.~A.~Thompson, A.~Burrows and P.~A.~Pinto,
Astrophys.\ J.\  {\bf 592}, 434 (2003)
[astro-ph/0211194].

\bibitem{Liebendoerfer:2002ny}
M.~Liebend\"orfer, A.~Mezzacappa, O.~E.~B.~Messer, G.~Martinez-Pinedo,
W.~R.~Hix and F.~K.~Thielemann, 
Nucl.\ Phys.\ A {\bf 719}, 144 (2003)
[astro-ph/0211329].


\bibitem{Bethe:1980gq}
H.~A.~Bethe, J.~H.~Applegate and G.~E.~Brown,
Astrophys.\ J.\  {\bf 241}, 343 (1980).

\bibitem{burrows_mazurek}
A.~Burrows and T.~J.~Mazurek,
Astrophys.\ J.\  {\bf 259}, 330 (1982).

\bibitem{Rampp:2002bq}
M.~Rampp and H.-T.~Janka,
Astron.\ Astrophys.\  {\bf 396}, 361 (2002)
[astro-ph/0203101].

\bibitem{Liebendoerfer:2003es}
M.~Liebend\"orfer, M.~Rampp, H.-T.~Janka and A.~Mezzacappa,
astro-ph/0310662.

\bibitem{progenitors}
K.~Nomoto and M.~Hashimoto,
Phys.\ Rep.\ {\bf 163}, 13 (1988) (Model n13); 
%
S.~E.~Woosley and T.~A.~Weaver,
Astrophys.\ J.\ Suppl.\  {\bf 101}, 181 (1995) (Model s15s7b2);
Woosley, S.E., Heger, A., and  Weaver, T.A. 2002, 
Rev.\ Mod.\ Phys.\ {\bf 74}, 1015 (2002) (Models s11.2, s20, s15a28, s25a28);
%
M.~Limongi, O.~Straniero, and A.~Chieffi, 
Astrophys.\ J.\ Suppl.\ {\bf 129}, 625 (2000) (Models l15, l25).


\bibitem{Liebendoerfer:2002ef}
M.~Liebend\"orfer, O.~E.~B.~Messer, A.~Mezzacappa, W.~R.~Hix,
F.~K.~Thielemann and K.~Langanke, 
``The importance of neutrino opacities for the accretion luminosity in
spherically symmetric supernova models,''
{\it Procs. 11th Workshop
on Nuclear Astrophysics}, Ringberg Castle, Tegernsee, Germany,
Feb.~11--16, 2002, Wolfgang Hillebrandt and Ewald M\"uller (Eds.); 
MPA-Report P13, MPI f\"ur Astrophysik, Garching
(2002) p.~126 [astro-ph/0203260].

\bibitem{Bruenn:1985en}
S.~W.~Bruenn,
Astrophys.\ J.\ Suppl.\  {\bf 58}, 771 (1985).

\bibitem{messerthesis}
O.~E.~B.~Messer, 
PhD Thesis, University of Tennessee (2000); 
Bulletin of the American Astronomical Society, Vol. 33, p.1411 (2001).

\bibitem{Langanke:2003ii}
K.~Langanke {\it et al.},
Phys.\ Rev.\ Lett.\  {\bf 90}, 241102 (2003)
[astro-ph/0302459].

\bibitem{Hix:2003fg}
W.~R.~Hix {\it et al.},
Phys.\ Rev.\ Lett.\  {\bf 91}, 201102 (2003)
[astro-ph/0310883].

\bibitem{Langanke:2001td}
K.~Langanke, G.~Martinez-Pinedo and J.~M.~Sampaio,
Phys.\ Rev.\ C {\bf 64}, 055801 (2001)
[nucl-th/0101039].

\bibitem{lsmodel}
J.~M.~Lattimer, C.~J.~Pethick, D.~G.~Ravenhall and D.~Q.~Lamb,
Nucl.\ Phys.\ A {\bf 432}, 646 (1985);
J.~M.~Lattimer and F.~D.~Swesty, 
Nucl.\ Phys.\ {\bf A535}, 331  (1991).

\bibitem{swestymira}
F.~D.~Swesty, J.~M.~Lattimer, and E.~S.~Myra,
Astrophys.\ J.\  {\bf 425}, 195 (1994).

\bibitem{marekdiploma}
A.~Marek, Diploma Thesis, Technische Universit\"at M\"unchen (TUM)
2003.

\bibitem{Janka:2004tt}
H.-T.~Janka, R.~Buras, F.~S.~Kitaura Joyanes, A.~Marek and M.~Rampp,
``Core-Collapse Supernovae: Modeling between Pragmatism and Perfectionism,''
in: Proceedings of {\it 12th Workshop on Nuclear Astrophysics},
Ringberg Castle, Tegernsee, March 22--27, 2004,
Eds. E.~M\"uller and H.-Th.~Janka, Report MPA-P14,
Max-Planck-Institut f\"ur Astrophysik, Garching, p.~150 (2004) 
[astro-ph/0405289].


\bibitem{shenmodel}
H.~Shen, H.~Toki, K.~Oyamatsu and K.~Sumiyoshi,
Nucl.\ Phys.\ A {\bf 637}, 435 (1998)
[nucl-th/9805035];
Prog. Theor. Phys. {\bf 100}, 1013 (1998) 
[nucl-th/9806095].


\bibitem{wolffmodel}
W.~Hillebrandt, R.~G.~Wolff and K.~Nomoto, 
Astron. Astrophys. {\bf 133} 175 (1984);
W.~Hillebrandt and R.~G.~Wolff, in {\it Nucleosynthesis:
 Challenges and New Developments}, W.~D.~Arnett and
J.~W.~Truran (Eds.), Univ. Chicago Press (1985), p.~131.


\bibitem{Burrows:2004va}
A.~Burrows, R.~Walder, C.~D.~Ott and E.~Livne,
``Rotating Core Collapse and Bipolar Supernova Explosions,''
in {\it The Fate of the Most Massive Stars},
Proc.~Eta Carinae Science Symposium, Jackson Hole,
May 23--28, 2004, ASP Conference Series (2005)
[astro-ph/0409035].

\bibitem{Heger:2004qp}
A.~Heger, S.~E.~Woosley and H.~C.~Spruit,
astro-ph/0409422.

\bibitem{Blondin:2002sm}
J.~M.~Blondin, A.~Mezzacappa and C.~DeMarino,
Astrophys.\ J.\  {\bf 584}, 971 (2003)
[astro-ph/0210634].

\bibitem{Scheck:2003rw}
L.~Scheck, T.~Plewa, H.-T.~Janka, K.~Kifonidis and E.~M\"uller,
Phys.\ Rev.\ Lett.\  {\bf 92}, 011103 (2004)
[astro-ph/0307352].


\bibitem{kotake}
K.~Kotake, S.~Yamada, and K.~Sato,
Astrophys.\ J.\  {\bf 618}, 474 (2005).


\bibitem{Fryer:1999he}
C.~L.~Fryer and A.~Heger,
Astrophys.\ J.\  {\bf 541}, 1033 (2000).

\bibitem{moenchmeyer}
H.~-T.~Janka and R. M\"onchmeyer, 
Astron.\ Astrophys.\ {\bf 209}, L5 (1989);
Astron.\ Astrophys.\ {\bf 226}, 69 (1989).

\bibitem{Ando:2003is}
S.~Ando and K.~Sato,
JCAP {\bf 0310}, 001 (2003)
[hep-ph/0309060].

\bibitem{Akhmedov:2003fu}
E.~K.~Akhmedov and T.~Fukuyama,
JCAP {\bf 0312} 007  (2003)
[hep-ph/0310119].

\bibitem{Ando:2004qe}
S.~Ando,
Phys.\ Rev.\ D {\bf 70} 033004 (2004)
[hep-ph/0405200].


\bibitem{Maltoni:2004ei}
M.~Maltoni, T.~Schwetz, M.~A.~T\'ortola and J.~W.~F.~Valle,
hep-ph/0405172.

\bibitem{sno}
\url{http://www.sno.phy.queensu.ca/}

\bibitem{Gil-Botella:2003sz}
I.~Gil-Botella and A.~Rubbia,
JCAP {\bf 0310}, 009  (2003)
[hep-ph/0307244]. 

\bibitem{distanceaccuracy} 
H.~Scheffler and H.~Els\"asser,
{\em Physics of the Galaxy and Interstellar Matter\/},
Springer-Verlag, Berlin 1988. 

\bibitem{pm}
P.~Mazzali, private communcation.


\bibitem{HK}
Y.~Itow {\it et al.},
arXiv:hep-ex/0106019.

\bibitem{uno}
The UNO whitepaper, ``Physics Potential and Feasibility of UNO,''
available at \url{http://ale.physics.sunysb.edu/uno/}.


\bibitem{Tomas:2003xn}
R.~Tom\`as, D.~Semikoz, G.~G.~Raffelt, M.~Kachelrie\ss~ and A.~S.~Dighe,
Phys.\ Rev.\ D {\bf 68}, 093013 (2003)
[hep-ph/0307050]. The neutral current events on oxygen have been
simulated following E.~Kolbe, K.~Langanke and P.~Vogel,
Phys.\ Rev.\ D {\bf 66}, 013007 (2002).


\bibitem{Beacom:2003nk}
J.~F.~Beacom and M.~R.~Vagins,
Phys.\ Rev.\ Lett.\  {\bf 93}, 171101 (2004)
[hep-ph/0309300].


\end{thebibliography}
\end{document}